\documentclass[letterpaper]{article}  
\usepackage{aaai25}   
\usepackage{times}   
\usepackage{helvet}   
\usepackage{courier}   
\usepackage[hyphens]{url}   
\usepackage{graphicx}  
\usepackage{natbib}    
\usepackage{caption}   
\usepackage{algorithm}
\usepackage{algorithmic}

\usepackage{amssymb}
\usepackage{amsmath}
\usepackage{bbding}
\usepackage{multirow}
\usepackage{subfigure}
\usepackage{booktabs}

\usepackage{newfloat}
\usepackage{listings}
\DeclareCaptionStyle{ruled}{labelfont=normalfont,labelsep=colon,strut=off} 
\floatstyle{ruled}
\newfloat{listing}{tb}{lst}{}
\floatname{listing}{Listing}

\usepackage{amssymb}
\usepackage{amsmath}

\title{Controllable Distortion-Perception Tradeoff Through Latent Diffusion for Neural Image Compression}
\author{
Chuqin Zhou\textsuperscript{\rm 1}, Guo Lu\textsuperscript{\rm 1}\thanks{Corresponding author}, Jiangchuan Li\textsuperscript{\rm 1}, Xiangyu Chen\textsuperscript{\rm 2}, Zhengxue Cheng\textsuperscript{\rm 1}, Li Song\textsuperscript{\rm 1}, Wenjun Zhang\textsuperscript{\rm 1}
}
\affiliations{

    Shanghai Jiao Tong University\textsuperscript{\rm 1}\\
    Institute of Artificial Intelligence (TeleAI), China Telecom\textsuperscript{\rm 2}\\
    \{zhouchuqin,luguo2014,cl4p-123,zxcheng,song\_li,zhangwenjun\}@sjtu.edu.cn,
    chxy95@gmail.com
    
}

\usepackage{bibentry}

\begin{document}

\maketitle
\begin{abstract}
Neural image compression often faces a challenging trade-off among rate, distortion and perception. While most existing methods typically focus on either achieving high pixel-level fidelity or optimizing for perceptual metrics, we propose a novel approach that simultaneously addresses both aspects for a fixed neural image codec. Specifically, we introduce a plug-and-play module at the decoder side that leverages a latent diffusion process to transform the decoded features, enhancing either low distortion or high perceptual quality without altering the original image compression codec. Our approach facilitates fusion of original and transformed features without additional training, enabling users to flexibly adjust the balance between distortion and perception during inference. Extensive experimental results demonstrate that our method significantly enhances the pretrained codecs with a wide, adjustable distortion-perception range while maintaining their original compression capabilities. For instance, we can achieve more than 150\% improvement in LPIPS-BDRate without sacrificing more than 1 dB in PSNR.
\end{abstract}

\section{Introduction}
As digital visual data continues to dominate Internet traffic, the development of efficient image and video codecs becomes increasingly crucial. In recent years, deep learning-based codecs have achieved significant advancements in both image domain~\cite{Cheng_2020_CVPR_DGML,Mentzer_2020_NeurIPS_HiFiC} and video domain~\cite{Hu_2021_CVPR_FVC,Li_2023_CVPR_DCVC_DC,Lu_2019_CVPR_DVC,Lu_21_PAMI_DVC,Lu_24_TCSVT_Preporcess}. These codecs have demonstrated a superior compression performance compared to traditional codecs~\cite{Bellard_2018_BPG,Bross_2021_TCSVT_VVC}. 

Current learning-based image codecs primarily rely on the transform coding paradigm and variational autoencoders (VAEs)~\cite{Balle_2018_ICLR_hyperprior}.  Most of these models use a rate-distortion loss function, directly optimizing for low distortion performance. However, distortion-oriented codecs often exhibit mode averaging behavior at low bitrates \cite{Zhao_2017_arXiv_VAE}, resulting in blurring that significantly degrades visual quality for human observers. 

\begin{figure}[t]
    \centering
    \includegraphics[width=1\linewidth]{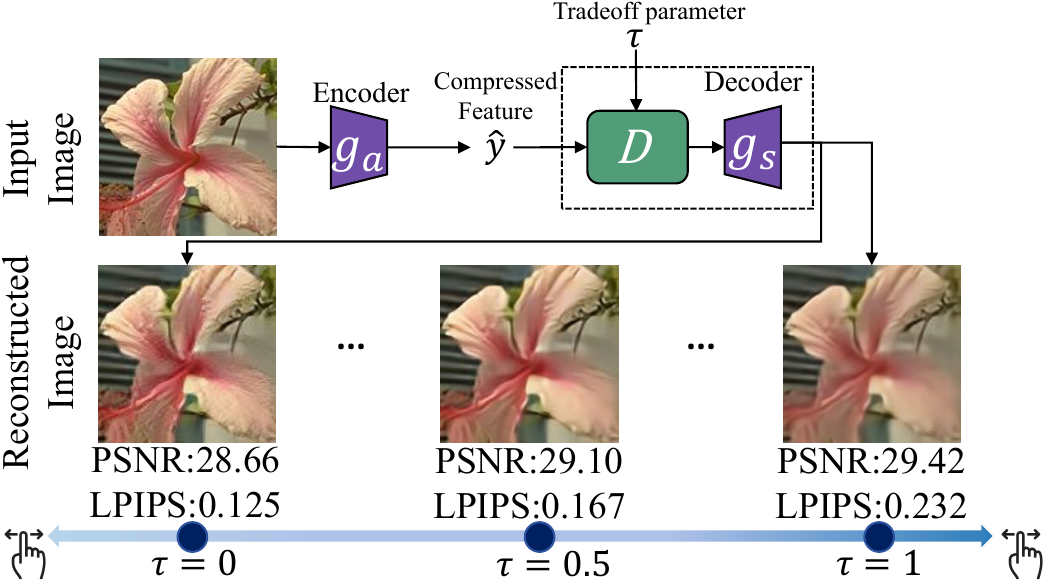}
    \caption{Overview of our proposed method. $\mathcal{D}$ represents a plug-and-play adaptive latent fusion module at decoder side for a base neural codec. We can achieve different distortion (PSNR) and perception (LPIPS) trade-offs, controlled by $\tau$. For simplicity, quantization and entropy coding are omitted.}
    \label{result}
\end{figure}

Recent studies demonstrate that optimizing for perceptual quality can lead to greater compression gains by allowing for imperceptible distortions, thereby reducing the bitrate. For example, HiFiC \cite{Mentzer_2020_NeurIPS_HiFiC} proposes to use Generative Adversarial Networks (GANs)~\cite{Goodfellow_2014_NeurIPS_GAN} to optimize neural image codecs. In contrast, CDC \cite{Yang_2023_NeurIPS_CDC} employs a diffusion-based \cite{Ho_2020_NeurIPS_DDPM} decoder to improve perceptual quality. However, both approaches struggle to achieve high pixel-level fidelity, as they may introduce high-frequency noise and unrealistic textures. \citet{Blau_2019_ICML_Rethinking} explains this phenomenon by highlighting a fundamental trade-off between perceptual quality and distortion. They suggest that these two goals cannot be fully achieved simultaneously within a given architecture.
We argue that both pixel-level fidelity and image-level realism are crucial for neural image compression frameworks. 
The inability to achieve both metrics simultaneously is a significant limitation in current image codecs. Given this limitation, an ideal codec should have the flexibility to traverse between different distortion-perception trade-offs at a given bitrate.

Recent research has explored flexible distortion-perception tradeoffs in image compression. A notable example is MRIC~\cite{Agustsson_2023_CVPR_MRIC}, which introduces a hyperparameter in the loss function to balance perception and distortion. This hyperparameter also serves as a condition for the decoder to adjust its reconstruction. However, this approach requires training the entire model from scratch and suffers from the inherent instability of GAN training. 

In this paper, we propose a novel compression pipeline that allows for a controllable trade-off between distortion and perception for a fixed pretrained codec, as illustrated in Fig. \ref{result}. Specifically, we introduce a plug-and-play adaptive latent fusion module at the decoder side, which transforms the decoded latent representations using a latent diffusion process \cite{Rombach_2022_CVPR_LDM}. This process allows representations originally optimized for low distortion to be converted to prioritize high perceptual quality, and vice versa. 

Assuming the base neural image codec is distortion-oriented, we first develop an auxiliary encoder, used only in the training stage, to generate guiding information optimized for perceptual quality. We then train the adaptive latent fusion module using perceptual loss while keeping the base codec's parameters fixed. In the inference stage, we fuse the original decoded feature with outputs from diffusion step based on user's preference, and the fused features are decoded by the original decoder. When integrated with existing variable bit rate schemes, our proposed model facilitates a trade-off among rate, distortion, and perception within a unified framework. Extensive experiments demonstrate the effectiveness of our framework. For distortion-oriented codecs, we achieve a more than 150\% improvement in LPIPS-BDRate with less than 1 dB sacrifice in PSNR.

Our main contributions are as follows:
\begin{itemize}
    \item We introduce an adaptive latent fusion module that enables controllable reconstruction at the decoder side, offering varying distortion-perception trade-offs.
    \item Our method serves as a plug-and-play module for fixed pretrained neural image codecs and is compatible with various compression frameworks. 
\end{itemize}

\section{Related Work}
\subsubsection{Image Compression.}
Image compression aims to reduce storage size by exploiting intra-image redundancy. Traditional image coding standards, such as JPEG~\cite{Wallace_1991_CACM_JPEG} and BPG~\cite{Sullivan_2012_TCSVT_HEVC}, employ manually designed modules like DCT to enhance compression performance, guided by the rate-distortion principle. In this context, distortion is typically measured using mean square error. 

Recently, VAE-based Neural Image Compression (NIC) methods~\cite{Minnen_2018_NeurIPS_Joint,Cheng_2020_CVPR_DGML,He_2021_CVPR_Checkerboard,Zhu_2022_ICLR_SwinTCharm,Zheng_2024_AAAI_RGBD} have experienced significant advancements, surpassing the current state-of-the-art traditional image codecs like VVC~\cite{Bross_2021_TCSVT_VVC}. 
One of the representative works is the hyperprior-based method~\cite{Balle_2018_ICLR_hyperprior}, which models latents using extracted hyperprior information for enhanced compression. Subsequent research has further improved compression performance through more sophisticated architecture~\cite{Cheng_2020_CVPR_DGML,Zhu_2022_ICLR_SwinTCharm} or entropy model~\cite{He_2021_CVPR_Checkerboard}. However, it is noteworthy that most existing works optimize codecs based on the rate-distortion strategy, potentially introducing blur artifacts at low bitrate settings. 

To address these issues and enhance the realism of compressed images, researchers have introduced the perception loss to optimize the image codecs. At this juncture, the majority of mature technologies in this field are GAN-based. For example, HiFiC \cite{Mentzer_2020_NeurIPS_HiFiC} achieves much more realistic results at low-bitrate settings. MS-ILLM \cite{Muckle_2023_ICML_MSILLM} introduces a non-binary discriminator, further enhancing the perceptual quality of reconstructed images. 

\subsubsection{Diffusion Probabilistic Models.}
Denoising diffusion probabilistic models~(DDPMs)~\cite{Ho_2020_NeurIPS_DDPM} generate data through a series of iterative stochastic denoising steps. The joint distribution of the data $x_0$ and the latent variable $x_{1:T}$ is learned through the model, i.e., $p_\theta(x_0)=\int p_\theta(x_{0:T})dx_{1:T}$. The goal of DDPMs is using network $\epsilon_\theta(x_t,t)$ to predict noise $\epsilon$ from a noisy image $x_t$ at a noise level $t$, where the noise $\epsilon$ is used to perturb a particular image $x_0$ through $x_t(x_0)=\sqrt{\alpha_t}x_0+\sqrt{1-\alpha_t}\epsilon$.

DDPMs have demonstrated success in various application areas, including image and video generation~\cite{Podell_2024_ICLR_SDXL,Lu_2024_ICLR_VDT}, super-resolution~\cite{Wang_2023_arxiv_StableSR,Luo_2024_AAAI_SkipDiff}, and restoration~\cite{Yu_2024_arxiv_SUPIR}. Recent research has extended their application to image compression. CDC~\cite{Yang_2023_NeurIPS_CDC} uses the diffusion model as a decoder, employing compressed features as conditions to guide the diffusion process in the image domain. HFD \cite{Hoogeboom_2023_arXiv_HFD} implements the diffusion model as an image restorer, using the low-quality decoded image to condition the diffusion process. Latent Diffusion Models (LDMs) proposed by \citet{Rombach_2022_CVPR_LDM} execute the diffusion process in latent domain of VAE. While \citet{Careil_2023_ICLR_Perco,Pan_2022_arxiv_extreme} have applied LDM paradigm to image compression, they all focus on ultra-low bitrates, where reconstruction preserves only semantic information, not pixel-level fidelity. We aim to explore LDM applications at a more general bitrate setting, seeking to maintain low distortion while achieving high perceptual quality.

Given the fundamental trade-off between perceptual quality and distortion, all of the aforementioned compression works focus on achieving either lower distortion or better perceptual quality. 
Approaches focusing on reducing distortion tend to result in blurring at low bitrates, while those prioritizing perceptual quality often introduce unrealistic noise.

\subsubsection{Distortion-Perception Trade-off in Compression.}

\begin{figure*}[t]
    \centering
    \includegraphics[width=1\linewidth]{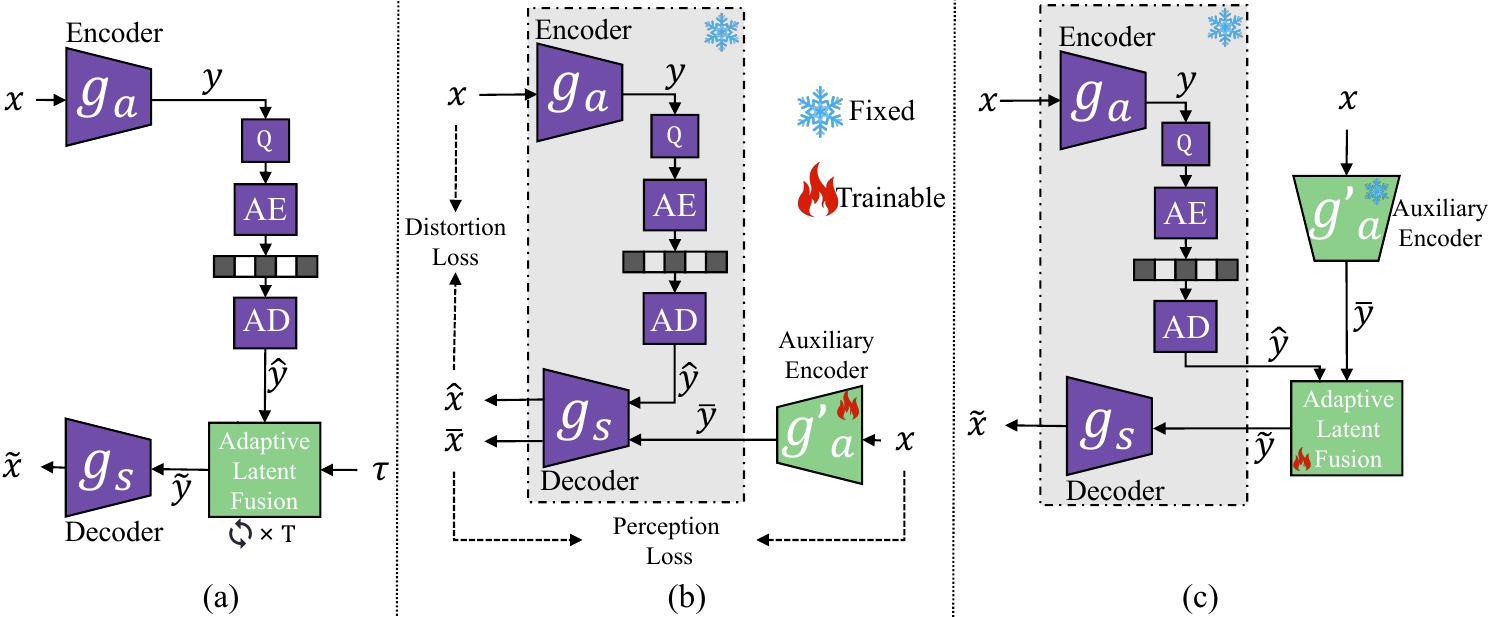}
    \caption{Illustration of the proposed method. For simplicity, we assume the base NIC is distortion-oriented. (a) represents the inference stage of our proposed pipeline. (b) and (c) represent the training procedures. We first train an auxiliary encoder $g'_a$ for the fixed base neural codec. Then, we train a plug-and-play adaptive latent fusion module to transform the original latent representations into features optimized for perceptual quality.}
    \label{overview}
\end{figure*}

Since it is challenging to optimize both simultaneously, there are attempts to allow users to choose between the two. \citet{Zhang_21_NeurIPS_RDB} propose achieving this trade-off through different decoders. \citet{Yan_22_ICML_control} introduce an approach using two decoders with interpolation in the image domain. MRIC \cite{Agustsson_2023_CVPR_MRIC} introduces a hyperparameter $\beta$ at training and inference stages to control the weight of distortion and perception. Building on this, \citet{Iwai_2024_WACV_CRDR} proposes controlling the quantization process with hyperparameters $q,\beta$ to achieve tunable rate, distortion, and realism. DIRAC \cite{Ghouse_2023_arXiv_DIRAC} adopts an approach similar to HFD \cite{Hoogeboom_2023_arXiv_HFD}, where the  output of the diffusion model is a residual added to the original low-quality image. This method can control the diffusion steps to achieve a distortion-perception trade-off.

However, these methods require training the entire model from scratch, limiting their ability to leverage the strengths of pretrained codecs. In contrast, we propose a plug-and-play module, akin to the plug-and-play characteristic of the mask sampling module in \citet{Liu_23_arxiv_Mask}, to address the limitations of the original codec without modifying it, thereby preserving its inherent advantages.

\section{Proposed Method}

\subsection{Overview}

The framework of our proposed method is shown in Fig.~\ref{overview} (a). To achieve a controllable distortion-perception trade-off, we introduce a plug-and-play adaptive latent fusion module on the decoder side of the existing pretrained codec. Our module can transform the original distortion-oriented features into perception-oriented features and vice versa. It also enables the fusion of these two types of features through weighted interpolation, resulting in decoded images with varying distortion-perception trade-offs. 

Our approach is flexible and can be applied to various baseline codecs, whether distortion-oriented or perceptual-oriented. Here we use a basic VAE codec as an example. Specifically, the decoded latent feature $\hat{y}$, obtained by arithmetic decoding (AD), serves as the diffusion condition. Guided by the distortion-perception trade-off parameter $\tau$, the adaptive latent fusion module generates a controllable feature $\tilde y$ with varying distortion-perception tradeoffs trough multiple diffusion steps. Finally, the fixed decoder $g_s$ converts $\tilde{y}$ into a decoded image $\tilde{x}$. Our pipeline enables controllable reconstruction without modifying the existing network architecture or retraining the model.

\subsection{Auxiliary Encoder for Baseline NIC}

While many studies have focused on optimizing image compression architectures, a classical VAE architecture is illustrated on Fig. \ref{overview} (b). The corresponding loss function is as follows:
\begin{equation}
    \min \mathcal{R}(Q(g_a(x)))+\beta\cdot \mathcal{L}(x,g_s(Q(g_a(x)))),
\end{equation}

\noindent where $Q$ denotes quantization and $\mathcal{R}(\cdot)$ represents the bitrate of the quantized latent representation. In distortion-oriented methods, $\mathcal{L}(x,\hat x) = ||x-\hat x||_2^2$ is used to measure distortion between the input and the reconstruction. For perception-oriented methods, $\mathcal{L}(x,\hat x) $ is defined based on the perceptual metrics such as LPIPS~\cite{Zhang_2018_CVPR_LPIPS}. 

For clarity, we assume that our base neural codec is distortion-oriented unless otherwise specified in this paper.
In the proposed framework, we aim to transform the existing quantized feature $\hat{y}$ to the controllable feature $\tilde{y}$ optimized for perceptual metrics. However, this transformation is non-trivial since the distributions of latent features optimizing for distortion or perception could be totally different, which is challenging even for the powerful diffusion methods. To address this issue, we further propose an auxiliary encoder to generate the corresponding guiding information, which is only used in the training stage.

Specifically, as shown on Fig. \ref{overview} (b), we introduce auxiliary encoder $g'_a$ with the same structure as the original encoder $g_a$, directly connected to the fixed original decoder $g_s$ during training. We optimize $g'_a$ by minimizing perceptual loss between input and reconstruction, such as LPIPS, while keeping all other modules frozen.
Then the optimized feature $\bar{y}$ will preserve more perceptual information and is employed as the auxiliary information for the training of our adaptive latent fusion module.  

\subsection{Adaptive Latent Fusion}

\begin{figure}[t]
    \centering
    \includegraphics[width=1\linewidth]{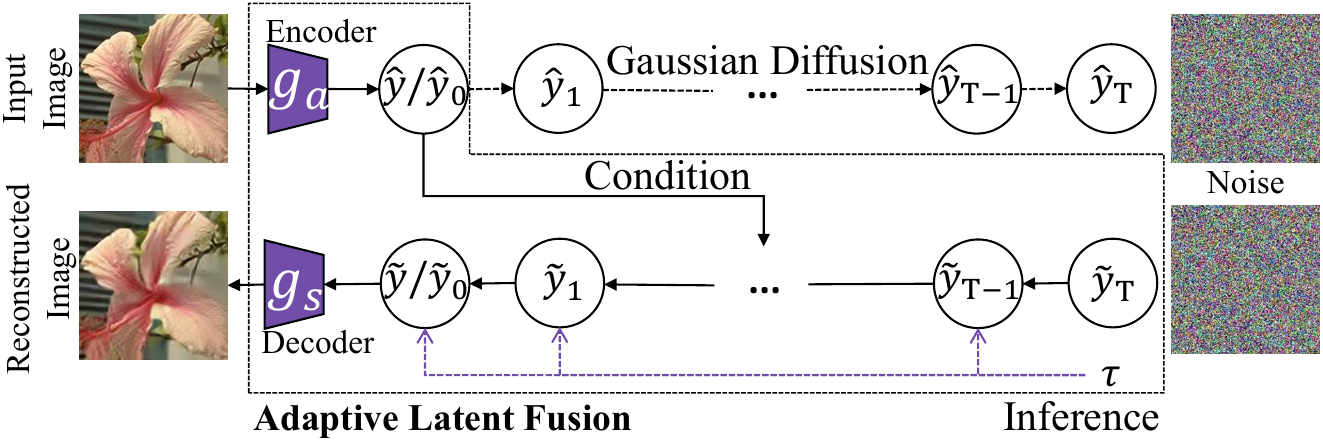}
    \caption{Overview of the latent diffusion process. For simplicity, we omit quantization and entropy coding modules. $\tau$ controls the diffusion process to achieve different tradeoffs.}
    \label{diffusion}
\end{figure}

Our proposed Adaptive Latent Fusion module transforms the decoded $\hat y$ into the controllable feature $\tilde y$, allowing for a desirable reconstruction that balances distortion and perception. The overview of this procedure is shown in Fig. \ref{diffusion}. The compressed latent $\hat y$ serves as a condition for the diffusion process and can be fused with the transformed feature during inference.

\begin{figure}[t]
    \centering
    \includegraphics[width=1\linewidth]{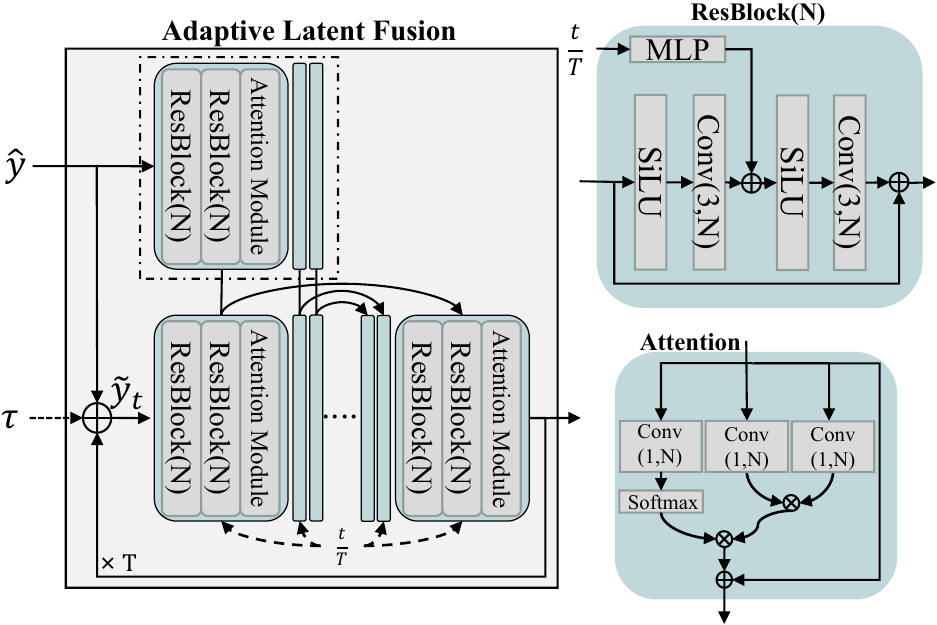}
    \caption{Architecture of our Adaptive Latent Fusion module. ResBlock(N) represents the ResBlock units with N channels. Conv(M,N) is a convolution layer with N channels, with $M\times M$ filters.}
    \label{architecture}
\end{figure}

The architecture of our network, dubbed $\mathcal{D}$, is shown in Fig. \ref{architecture}. We employ the classical latent diffusion architecture, which consists of M units, each containing two time-aware ResNet blocks and an attention block. To utilize the conditional information like the original feature $\hat y$ for the diffusion procedure, we generate condition information using the same unit and perform conditioning by concatenation. In addition, given the complexity of predicting noise for features with large channels and the strong condition $\hat y$, we directly learn transformed features $\tilde y$ instead of the noise, based on the original feature $\hat y$ and the pseudo-continuous variable $\frac{t}{T}$. 
More importantly, we further use the hyper-parameter $\tau$ to control the diffusion procedure and produce the features with varying distortion-perception trade-offs. 

\subsubsection{Training.}
As shown in Fig. \ref{overview} (c), we use the auxiliary encoder to generate auxiliary information $\bar y$ as the training target.
During training, we add noise to the decoded feature $\hat y$. Thus, the diffusion input is given by $\hat y_t=\sqrt{\alpha_t}\hat y+\sqrt{1-\alpha_t}\epsilon$, where $\alpha_t=\prod_{s =1}^t (1-\beta_s)$ and $\beta_t\in(0,1)$ is a monotonically increasing sequence of noise scheduler. $t$ is randomly sampled from $[0,T]$ during training. We use the pseudo-continuous variable $\frac{t}{T}$ to indicate noise intensity to the model. This allows us to use an arbitrary and fewer number of denoising steps during inference.

We disregard noise level $t$ and directly input the result of $\mathcal D$ at each step into the decoder $g_s$. The LPIPS distance between the reconstructed image and the original image is included in the loss function. The training loss function for adaptive latent fusion module is formulated as follows:
\begin{equation}
    \min \lambda||\bar y - \mathcal{D}(\hat y_t,\hat y, \frac tT)||_2^2+\mathcal{L}(x,g_s(\mathcal{D}(\hat y_t,\hat y, \frac tT))),
\end{equation}
where the first term represents the loss for the diffusion procedure and the second term represents the reconstruction loss in image compression. $\lambda$ is a trade-off parameter for different losses. 

\subsubsection{Inference.}

During inference, we follow DDIM~\cite{Song_2021_ICLR_DDIM}, which replaces the original Markov process with a deterministic generative process to improve sampling speed. Our method aims to generate a decodable feature directly and calculate the predicted noise. As shown in Fig. \ref{diffusion}, to differentiate from the training process, we refer to $\tilde y_t$ as the noisy feature during inference. At timestep $t$ given the predicted result $\mathcal{D}(\tilde y_t,\hat y,\frac tT)$, we can derive the equivalence with $\epsilon(\tilde y_t,\hat y,\frac tT)=\frac{\tilde y_t-\sqrt{\alpha_t}\mathcal{D}(\tilde y_t,\hat y,\frac tT)}{\sqrt{1-\alpha_t}}$, which is the predicted noise. The sampling process can be formulated as:
\begin{equation} 
    \tilde y_{t-1}=\sqrt{\alpha_{t-1}}\mathcal D(\tilde y_t,\hat y, \frac tT)+\sqrt{1-\alpha_{t-1}}\epsilon(\tilde y_t, \hat y, \frac tT),
\end{equation}

We can use the above recurrence equation to sample $\tilde y_{t-1}$ from $\tilde y_t$ as $t$ gradually decreases from $T$ to 0, while $\tilde y_T\sim \mathcal{N}(\mathbf{0},\mathbf{I})$. It should be noted that we use different values for $T$ in the training and inference process, which significantly reduces the number of diffusion steps required for inference.

In practice, different scenarios have different requirements for distortion and perception. Therefore, we aim to achieve a controllable trade-off between them during the inference phase. We use the following equation to guide the sampling process and apply $\tau$ for weighted interpolation between the predicted output $\tilde y_{t-1}$ and the original latent $\hat y$.
\begin{align}
    \tilde y_{t-1}&=\sqrt{\alpha_{t-1}}[(1-\tau^2)\times\mathcal{D}(\tilde y_t,\hat y,\frac tT)+\tau^2\times \hat y]\nonumber\\
    &+(1-\tau^2)\times\sqrt{1-\alpha_{t-1}}\epsilon(\tilde y_t,\hat y,\frac tT),
\end{align}

\begin{figure*}[t]
    \centering
    \includegraphics[width=1\linewidth]{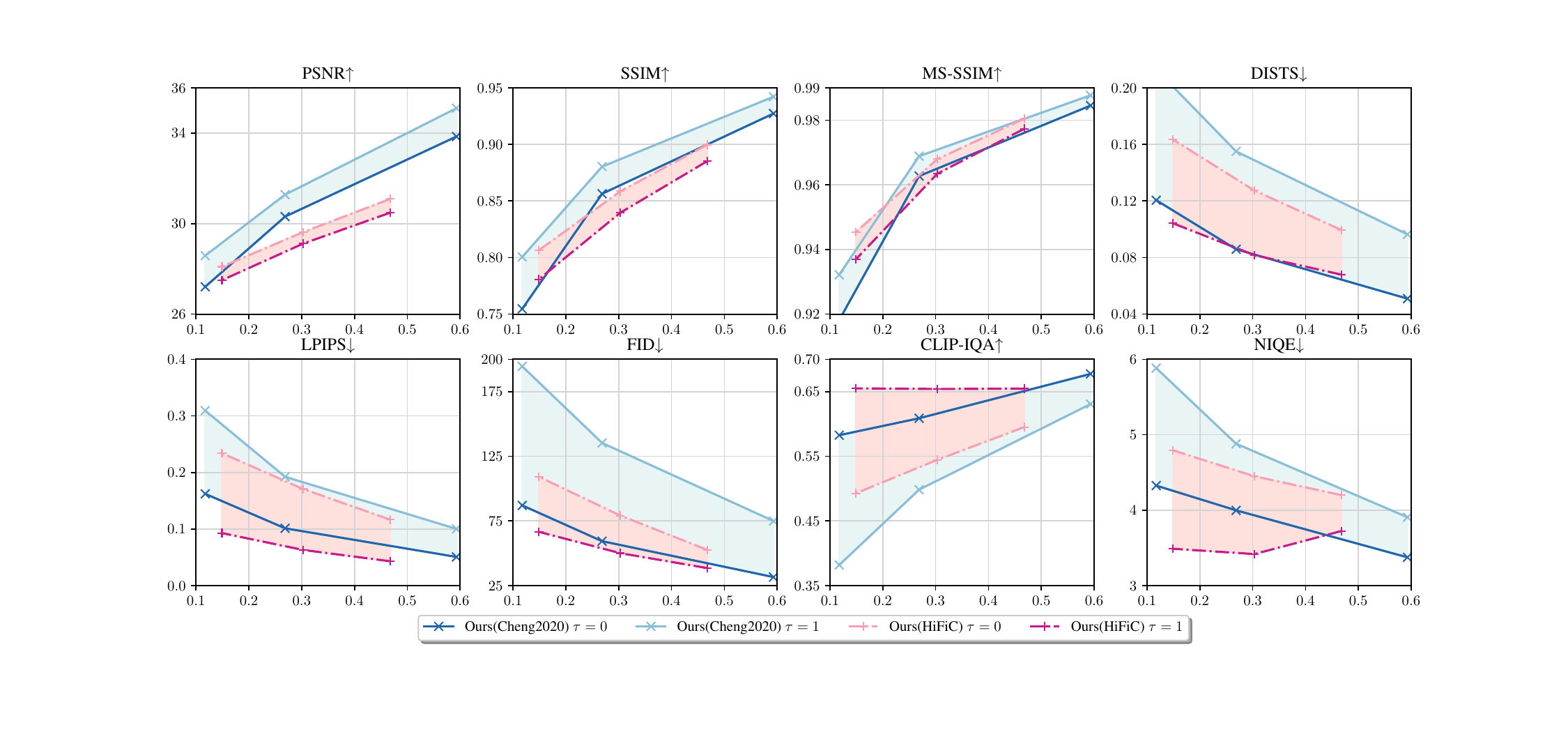}
    \caption{Trade-offs between bitrate  and different metrics  for various base codecs tested on Kodak dataset. Arrows in the plot titles indicate whether high($\uparrow$) or low($\downarrow$) values indices a better score.}
    \label{codec}
\end{figure*}

We use $\tau^2$ instead of $\tau$ to ensure that the two latent representations are weighted appropriately, achieving a more linear control effect. The parameter $\tau$ ranges from [0, 1], where $\tau=0$ results in outputs composed entirely of perception-oriented latents. Conversely, When $\tau=1$, $\tilde y_0=\hat y$, which corresponds to the base codec. The final decoded image $\tilde{x}$ is obtained by inputting the last $\tilde y_0$ into the decoder $g_s$.

\section{Experiments}

\subsection{Experimental Settings}
\subsubsection{Training Details.}

For training, we use a high-quality Flickr2W dataset~\cite{Liu_2020_arXiv_Flickr}, and randomly crop images to a resolution of $256\times 256$. To train the auxiliary encoder and the adaptive latent fusion module, we use the AdamW optimizer with a batch size of 32. The learning rate is maintained at a fixed value of $5\times 10^{-5}$. 

\subsubsection{Evaluation.}

\begin{figure*}[t]
    \centering
    \includegraphics[width=1\linewidth]{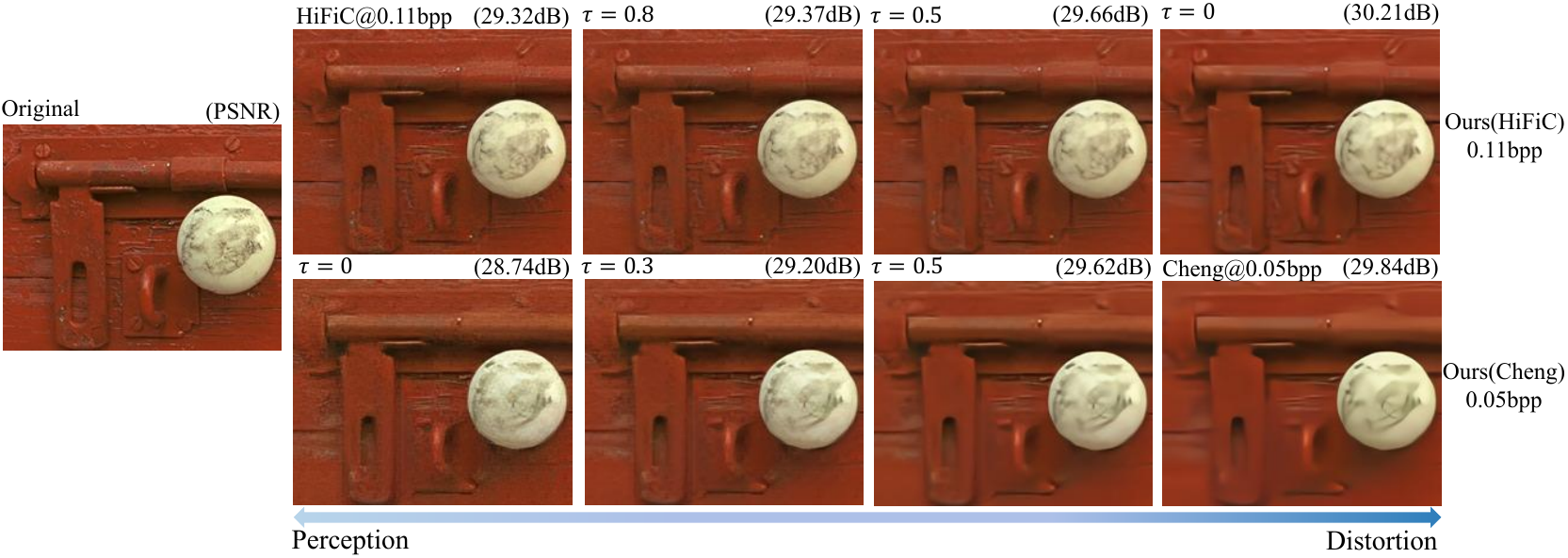}
    \caption{Kodak reconstructions of our method for different rate-distortion-perception. Shown scores are for full image.}
    \label{quantitative}
\end{figure*}

We evaluate our method using both distortion metrics and perceptual quality metrics. All evaluations are performed on full-resolution images. We select eight widely used metrics for image quality evaluation. To measure distortion and perception, we use PSNR, SSIM, DISTS, LPIPS~\cite{Zhang_2018_CVPR_LPIPS}, FID~\cite{Heusel_2017_NeurIPS_FID}, CLIP-IQA~\cite{Wang_2023_AAAI_CLIPIQA}, and NIQE~\cite{Mittal_2013_SPL_NIQE}. Among these metrics, FID, CLIP-IQA, NIQE are non-reference metrics, while others are full-reference metrics. When calculating FID, we follow the procedure of previous compression works~\cite{Mentzer_2020_NeurIPS_HiFiC} by segmenting images into non-overlapping patches of $256\times 256$ resolution. The evaluations are conducted on two common image compression benchmark datasets: the CLIC2020 test set and the Kodak dataset.

\subsubsection{State-of-the-art Methods.}
We compare our method with several representative neural compression approaches. MSHyper \cite{Balle_2018_ICLR_hyperprior} introduces the hyperprior for enhanced compression. Cheng2020 \cite{Cheng_2020_CVPR_DGML} employs an attention mechanism and outperforms the traditional VVC codec. HiFiC \cite{Mentzer_2020_NeurIPS_HiFiC}, a GAN-based codec trained for specific rate-perception trade-offs, exemplifies a leading perceptual codec. CDC \cite{Yang_2023_NeurIPS_CDC} uses a conditional DDIM decoder to generate reconstructions from latent representations, providing distortion-oriented (CDC) and perception-oriented (CDC-lpips) models. MRIC \cite{Agustsson_2023_CVPR_MRIC} introduces $\beta$ to achieve various distortion-perception tradeoffs. We limit comparisons to studies with publicly available codes and models for consistent testing and evaluation.

\subsection{Main Results}

\subsubsection{Distortion-Perception Trade-off Ability.}

As a plug-and-play approach, our approach can be easily integrated with the existing image codecs. We select two representative methods, Cheng et al. \cite{Cheng_2020_CVPR_DGML}(denoted as Ours(Cheng2020)) and HiFiC \cite{Mentzer_2020_NeurIPS_HiFiC}(denoted as Ours(HiFiC)) as our base codecs and evaluate the versatility and effectiveness of the proposed approach.  More test results for other base models can be found in the Appendix.

In Fig. \ref{codec}, the shaded area represents the adjustable rate-distortion-perception range achievable by a single model. Our model is presented in two configurations: $\tau\in\{0,1\}$. When $\tau=1$, our codec is equivalent to the base codec. When $\tau=0$, the optimization direction of our loss is opposite to the original codec. It is noted that our proposed approach can achieve a wide range of trade-offs between distortion and perception in different bitrates. 
In addition, our pipeline preserves the original codec and this enables us to maintain state-of-the-art (SOTA) performance in areas where the original codec excels. 

For the distortion-optimized base codec, our approach, Ours (Cheng2020, $\tau=0$), achieves a significant improvement over the original Cheng's method, enhancing LPIPS-BDRate by 158.75\%, which corresponds to an average LPIPS gain of 0.096, with only a modest PSNR degradation of 1.08 dB on average. On the high-perception side ($\tau=0$), our method matches or surpasses the state-of-the-art generative approach HiFiC in DISTS, while also delivering a substantial average PSNR improvement of 1.48 dB.

For the perception-oriented base codec, compared with the original HiFiC approach, our approach Ours(HiFiC, $\tau=0$) saves 22.59\% bitrate in terms of PSNR performance on Kodak dataset. On the low-distortion side ($\tau=0$), we match or outperform the state-of-the-art distortion method Cheng2020 in SSIM and MS-SSIM, while also significantly outperforming it in perceptual quality.

\subsubsection{Comparison with SOTA Methods.}

\begin{figure}[t]
    \centering
    \includegraphics[width=1\linewidth]{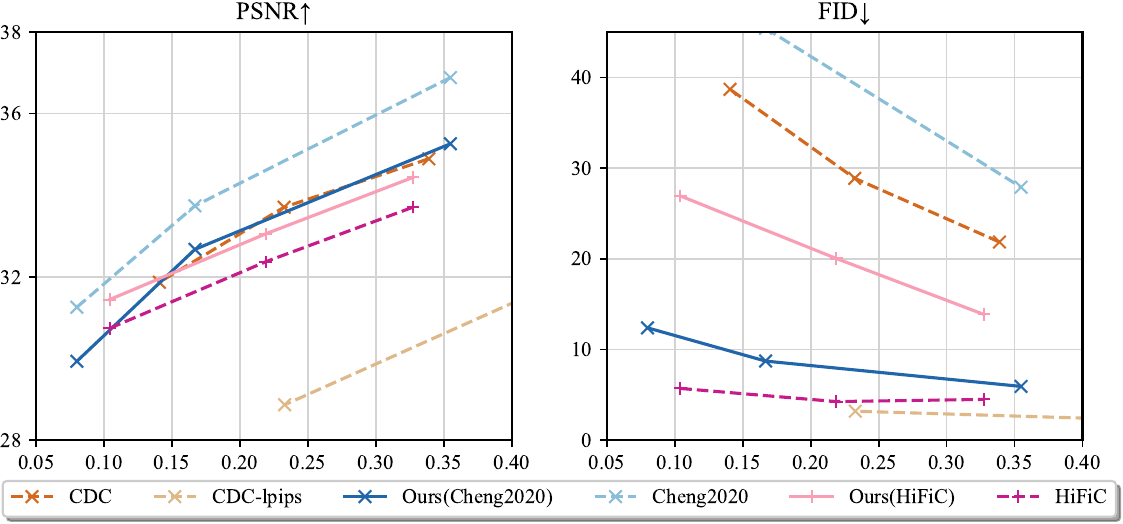}
    \caption{Trade-offs between bitrate and different metrics for various models tested on CLIC2020 test set. Ours are shown with $\tau=0$.}
    \label{clic}
\end{figure}

Fig. \ref{clic} provides more results when comparing our approach ($\tau=0$) with existing image codecs in the CLIC2020 dataset.

When compared with CDC method, which also employs a diffusion model, Ours(HiFiC, $\tau=0$) achieves a comparable distortion performance, while significantly enhancing perceptual quality. Ours(Cheng2020, $\tau=0$) significantly outperforms CDC-lpips in terms of PSNR while achieving a similar LPIPS score. Additionally, it surpasses CDC in LPIPS and FID while maintaining a similar PSNR. Despite not achieving the most SOTA performance in perceptual metrics due to the fixed decoder and the inherent limitations of encoder-extracted features, we observe substantial enhancement in perceptual quality and achieve comparable results with perception-oriented models, which is the main purpose of our design. In fact, given such a low LPIPS, further reductions may not yield perceptually significant improvements but could potentially introduce more high-frequency noise at low bit rates.

\subsubsection{Rate-Distortion-Perception Trade-off.}

\begin{figure}[t]
    \centering
    \includegraphics[width=1\linewidth]{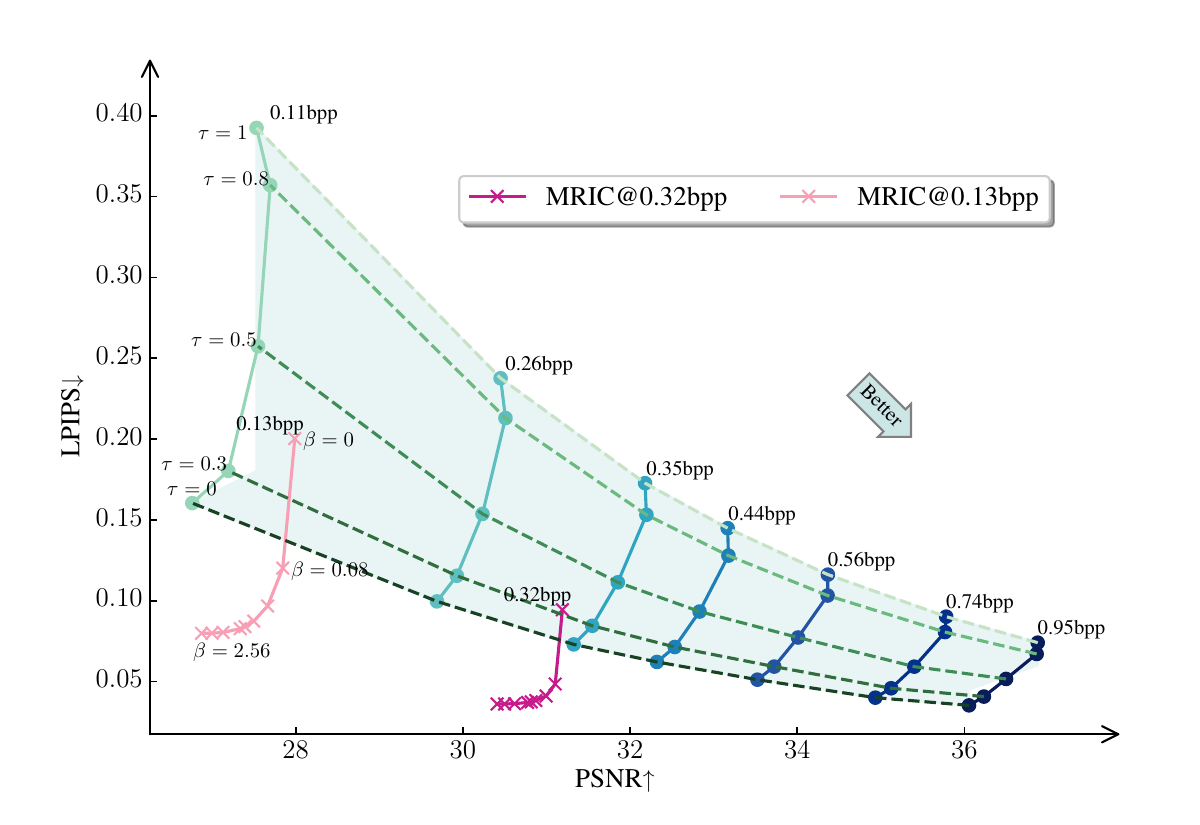}
    \caption{Distortion (PSNR) vs. perception(LPIPS) on Kodak for different rate-distortion-perception tradeoffs. 
    }
    \label{tradeoff}
\end{figure}

Our method enables the transformation of distortion and perception-oriented latent representations, allowing for a flexible tradeoff between these dimensions during the inference phase. By combining this approach with a variable bitrate scheme, we can achieve a three-dimensional exploration of the rate-distortion-perception tradeoff. Here we apply the variable bitrate method proposed by \cite{Cui_2021_CVPR_CVR} to the baseline method \cite{Balle_2018_ICLR_hyperprior}, denoted as Ours(MSHyper).

As illustrated in Fig. \ref{tradeoff}, we evaluate the effect of different $\tau$ values ($\tau\in \{0, 0.3, 0.5, 0.8, 1\}$) at various bit rates. Our results Ours(MSHyper) demonstrate that the proposed approach performs a controllable and smooth traversal between low distortion (high PSNR) and high perceptual quality (low LPIPS). MRIC \cite{Agustsson_2023_CVPR_MRIC}, on the other hand, is not a variable bitrate scheme and involves training the entire model from scratch, resulting in a non-linear and less controllable tradeoff between distortion and perception. Their method shows little or no change from $\beta=0.64$ to $\beta=2.56$, making their methods less controllable. Our approach allows for a linear transition between distortion and perception. Our method can achieve a much larger range of conversion, significantly improving perceptual quality, as evidenced by a 170\% improvement in LPIPS-BDRate.

\begin{figure}[t]
    \centering
    \includegraphics[width=1\linewidth]{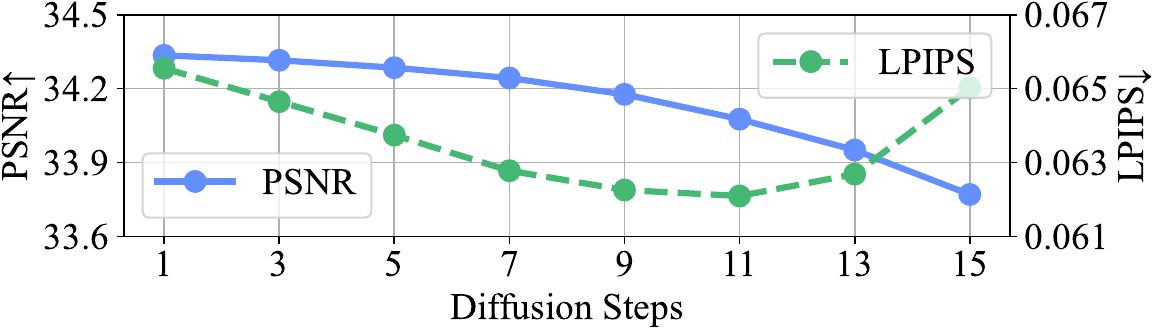}
    \caption{Reconstruction quality in the terms of PSNR and LPIPS versus the number of sampling timesteps.}
    \label{step}
\end{figure}

\subsubsection{Quantitative Results.}

The quantitative results are presented in Fig. \ref{quantitative}.
When $\tau=1$, our codec is equivalent to the base codec. Ours(Cheng, $\tau=0$) generates reconstructed image with comparable perceptual quality to HiFiC, achieving much higher PSNR even at a significantly lower BPP. Decreasing $\tau$ results in a drastic change in perceptual quality. Ours(HiFiC, $\tau=0$) increases PSNR while maintaining more texture details, avoiding the smoothness and blurriness that Cheng exhibits. Decreasing $\tau$ slightly reduces reconstruction texture while boosting PSNR. 

\subsection{Ablation Study}
\subsubsection{Effectiveness of Latent Diffusion Module.}

As shown in Table \ref{ablation}, we construct and evaluate different variants of our design. Negative values indicate the amount of bitrate saved compared to our method.

\begin{figure}[t]
    \centering
    \includegraphics[width=1\linewidth]{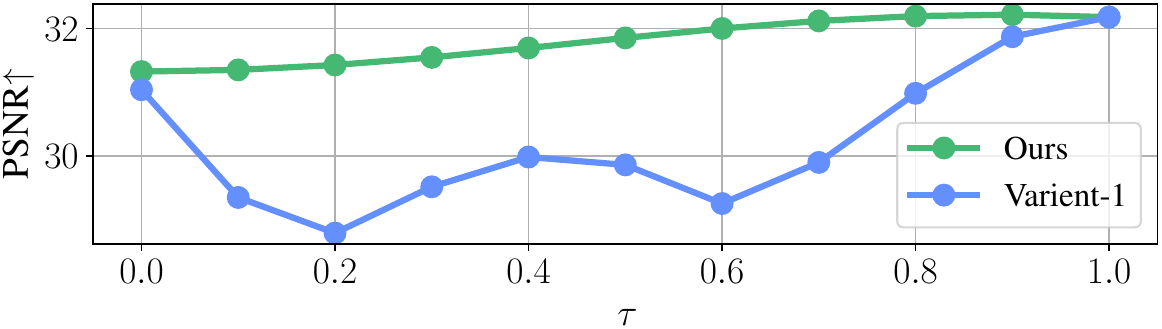}
    \caption{Comparison of PSNR values with varying hyper-parameters $\tau$ across different methods.}
    \label{fail}
\end{figure}

Variant-1 excludes the diffusion process, omitting the auxiliary encoder and its associated loss. Our method outperforms Variant-1 in both PSNR and LPIPS, demonstrating the effectiveness of the latent diffusion process and auxiliary encoder. As shown in Fig. \ref{fail}, Variant-1's features reveal a critical issue: PSNR fails to correlate positively with $\tau$ when fused with untreated features. While this design can decode the original features and produce modified features, it fails to achieve effective fusion of these two through simple weighted interpolation. 

To obtain Variant-2, the decoder is unfrozen and trained jointly. While Variant-2 achieves results similar to ours in terms of LPIPS and better PSNR, the decoder cannot handle unprocessed latents from the encoder as it has been modified. Consequently, Variant-2 does not offer the same flexibility in distortion-perception trade-offs as our method.

\begin{table}[t]
    \begin{center}
        \caption{Ablation study on different variants of our model on Kodak dataset. }
        \label{ablation}
        \setlength{\tabcolsep}{1mm}
        \begin{tabular}{c|cc|cc}
            \multicolumn{1}{l|}{\multirow{2}{*}{Methods}} & \multicolumn{1}{c|}{\multirow{2}{*}{Diffusion Process}} & \multirow{2}{*}{Decoder} & \multicolumn{2}{c}{BDRate} \\ 
            \multicolumn{1}{l|}{} & \multicolumn{1}{l|}{} & & PSNR & LPIPS \\ \hline
            Ours & \Checkmark & \XSolidBrush & 0 & 0\\
            Variant-1 & \XSolidBrush & \XSolidBrush & +6.91\% & +4.69\%\\
            Variant-2 & \XSolidBrush & \Checkmark & -2.36\% & +0.63\%\\
        \end{tabular}
    \end{center}
\end{table}

\subsubsection{Generation Speed.}

We investigate the impact of sampling timesteps on the reconstruction quality. The trade-off between generation speed and quality is illustrated in Fig. \ref{step}. Our method achieves competitive performance with just 10 timesteps, adopted as the default setting. However, exceeding 10 timesteps degrades LPIPS scores, as the diffusion module introduces excessive high-frequency components, causing a drift from the original content. In comparison, CDC \cite{Yang_2023_NeurIPS_CDC} employs DDIM in the pixel domain with 17 timesteps for direct image output.

\section{Conclusion}
In this paper, we present a plug-and-play method for neural image codecs, which allows for various distortion-perception reconstruction tradeoffs from a single latent representation. An adaptive latent fusion module can either transform the feature to a high-perception or a low-distortion one. We keep the image codecs unchanged, but allow the trade-off between realism and distortion to happen on the receiver side, with no change in the bit stream. By integrating with the variable bitrate codecs, users can select the desired rate and switch between reconstructions which are as close to the original as possible and those with a better level of detail. Experiments demonstrate that our method works effectively with both distortion-driven and perception-driven models, achieving remarkable performance improvements in areas where the original codec is less effective, while minimally sacrificing their respective advantages.

\section*{Acknowledgments}
This work is supported by the National Key Research and Development Program of China under Grant 2024YFF0509700,  National Natural Science Foundation of China(62471290,62331014) and the Fundamental Research Funds for the Central Universities.

\bibliography{ref}

\clearpage
\appendix
\onecolumn
\setcounter{figure}{0}
\setcounter{table}{0}
\section{Appendix / supplemental material}

\subsection{Pretrained Baselines}
We refer to \cite{Begaint_2020_arXiv_compressai} for pretrained Cheng2020 \cite{Cheng_2020_CVPR_DGML} model. For HiFiC \cite{Mentzer_2020_NeurIPS_HiFiC} and MRIC \cite{Agustsson_2023_CVPR_MRIC} model, we use the pretrained models implemented in the publicly available repositories\footnote{\url{https://github.com/Justin-Tan/high-fidelity-generative-compression}}\footnote{\url{https://github.com/Nikolai10/MRIC}}. For CDC \cite{Yang_2023_NeurIPS_CDC} baseline, we use the official codebase\footnote{\url{https://github.com/buggyyang/CDC_compression}}. 

\subsection{Denoising Diffusion Models}

This section briefly describes the denoising diffusion models covered in this paper. Denoising Diffusion Probability Models (DDPMs)\cite{Ho_2020_NeurIPS_DDPM} generates data through a series of iterative stochastic denoising steps. The joint distribution of the data $x_0$ and the latent variable $x_{1:T}$ is learned through the model, i.e., $p_\theta(x_0)=\int p_\theta(x_{0:T})dx_{1:T}$. While the diffusion process (denoted by $q$) gradually destroys the structure of the data, its reverse process $p_\theta$ gradually generates the structure. Both processes involve Markov dynamics in a series of transition steps, where a monotonically increasing sequence of noise variances $\beta_t\in(0,1)$ controls the diffusion process. The denoising process predicts the posterior mean from the diffusion process and is parameterized by a neural network $\epsilon_\theta(x_t,t)$. The goal of DDPMs is using network $\epsilon_\theta(x_t,t)$ to predict noise $\epsilon$ from a noisy image $x_t$ at a noise level $t$, where the noise $\epsilon$ is used to perturb a particular image $x_0$:
\begin{equation}
    q(x_t|x_{t-1})=\mathcal{N}(x_t|\sqrt{\alpha_t}x_{t-1},\beta_t\mathbf{I}),
\end{equation}
\begin{equation}
    p_\theta(x_{t-1}|x_t)=\mathcal{N}(x_{t-1}|\epsilon_\theta(x_t,t),\beta_t\mathbf{I}),
\end{equation}

\noindent where $t\sim Unif\{1,...,T\}$, $\epsilon\sim \mathcal{N}(\mathbf{0},\mathbf{I})$, $x_t(x_0)=\sqrt{\bar{\alpha}_t}x_0+\sqrt{1-\bar{\alpha}_t}\epsilon$, $\alpha_t=1-\beta_t$, $\bar{\alpha}_t=\prod_{s =1}^t (1-\beta_s)$. 

During testing, the data is generated using the sampling process of Langevin dynamics. The sampling process can be described by the following equation, which iteratively subtracts the predicted noise from intermediate noisy results to obtain the final predicted data.
\begin{equation}
    x_{t-1}=\frac1{\sqrt{\alpha_t}}\left(x_t-\frac{1-\alpha_t}{\sqrt{1-\bar{\alpha_t}}}\mathbf{\epsilon}_\theta(x_t,t)\right)+\sigma_t\epsilon,
\end{equation}

\noindent where $\sigma_t$ is a hyperparameter used to control the degree of randomness of the sampling process.

In addition, \cite{Song_2021_ICLR_DDIM} propose the Denoising Diffusion Implicit Models (DDIMs), which replaces the original Markov process with a deterministic generative process. By randomly sampling the initial data from the prior and guiding it directly to the final prediction $x_0$ at each step. DDIM significantly improves sampling speed. The sampling process is given by the following equation:
\begin{equation} 
    x_{t-1}=\sqrt{\bar{\alpha}_{t-1}}\underbrace{\frac{x_t-\sqrt{1-\bar{\alpha}_t}\epsilon_\theta(x_t)}{\sqrt{\bar{\alpha}_t}}}_{\text{"predicted }x_0\text{"}}+\underbrace{\sqrt{1-\bar{\alpha}_{t-1}}\epsilon_\theta(x_t)}_{\text{"direction pointing to }x_t\text{"}},
\end{equation}

For simplicity, we use $\alpha_t$ instead of $\bar{\alpha}_t$ in the article.

\begin{figure*}[ht]
    \centering
    \includegraphics[width=1\linewidth]{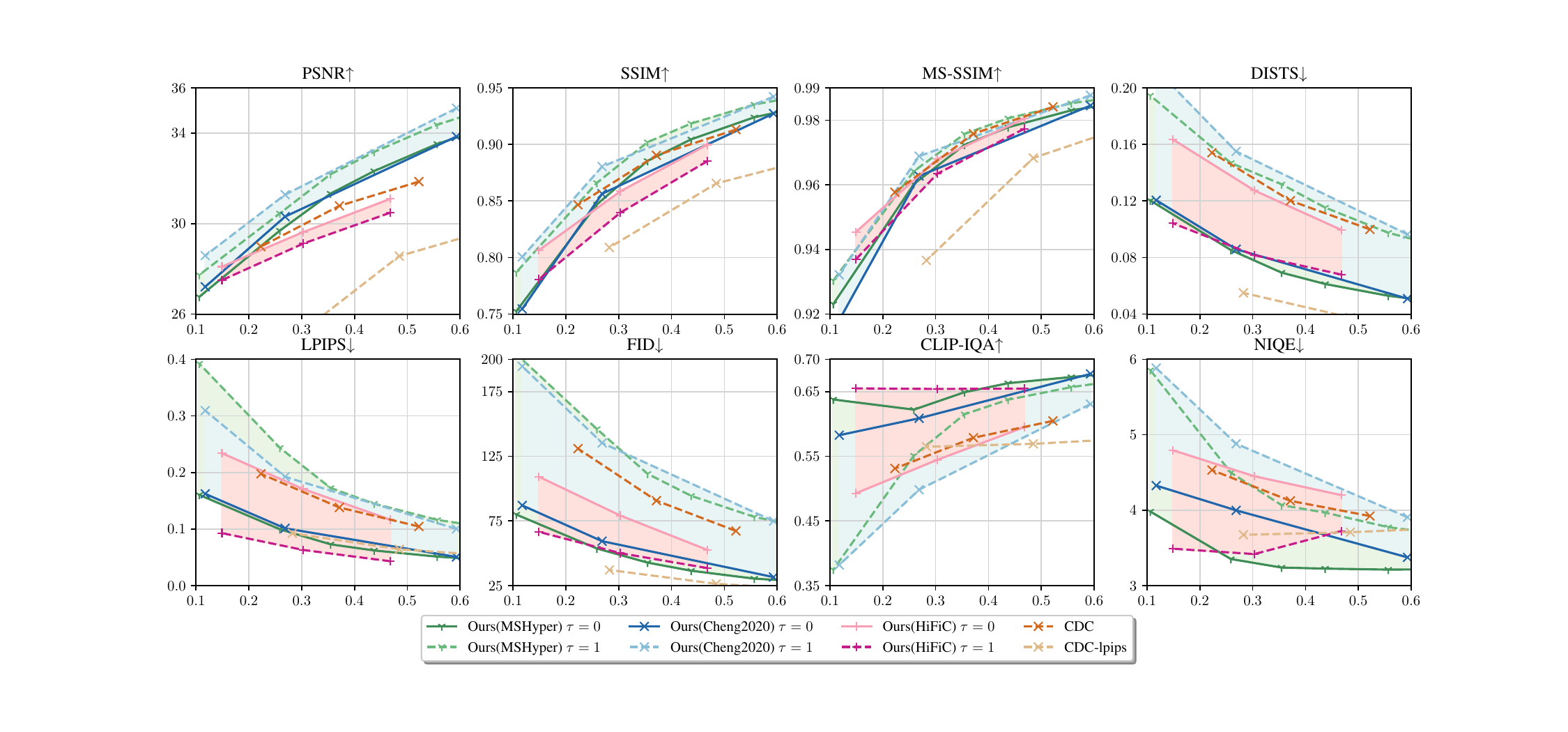}
    \caption{Trade-offs between bitrate  and different metrics  for various models tested on Kodak dataset. Arrows in the plot titles indicate whether high($\uparrow$) or low($\downarrow$) values indices a better score.}
    \label{kodak}
\end{figure*}

\subsection{Latent Diffusion Models}

This section briefly introduce the Latent Diffusion Models (LDM) proposed by \cite{Rombach_2022_CVPR_LDM}. LDM perform the diffusion process in the latent domain of a VQVAE. Specifically, the input image $x$ is first encoded to a latent representation $y=\mathcal{E}(x)$ using the encoder $\mathcal{E} $. The diffusion process learns the distribution of $y$ during training, which is then used to generate $\tilde y$ during testing. The generated $\tilde y$ is decoded back to the image domain through the decoder $\mathcal{D}$, yielding $\tilde x=\mathcal{D} (\tilde y)$.

\subsection{Additional Rate-Distortion-Perception Results}

More test results for different base codec are shown in this section.

As a plug-and-play approach, our approach can be easily integrated with the existing image codecs. We deploy our method with three representative methods, MSHyper \cite{Minnen_2018_NeurIPS_Joint}(denoted as Ours(MSHyper)), Cheng et al. \cite{Cheng_2020_CVPR_DGML}(denoted as Ours(Cheng2020)) and HiFiC \cite{Mentzer_2020_NeurIPS_HiFiC}(denoted as Ours(HiFiC)) as our base codecs and evaluate the versatility and effectiveness of the proposed approach. The results is shown in Fig. \ref{kodak}.

Our method enables the transformation of distortion and perception-oriented latent representations, allowing for a flexible tradeoff between these dimensions during the inference phase. As illustrated in Fig. 2, we evaluate the effect of different $\tau$ values ($\tau\in \{0, 0.3, 0.5, 0.8, 1\}$) at various bit rates for Ours(Cheng2020) and Ours(HiFiC). 

\begin{figure*}[htbp]
    \centering
    \subfigure[Ours(Cheng2020)]{
    \begin{minipage}[t]{0.49\linewidth}
    \centering
    \includegraphics[width=1\linewidth]{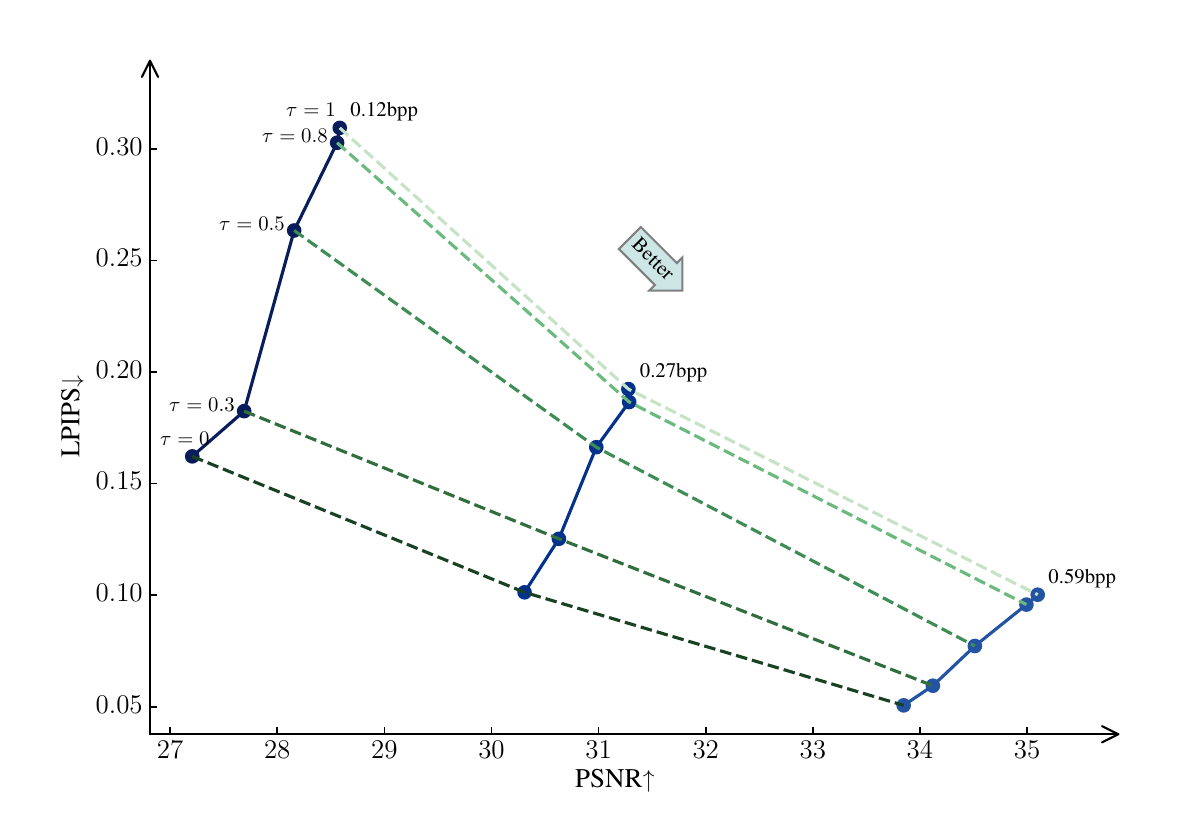}
    \end{minipage}%
    }%
    \subfigure[Ours(HiFiC)]{
    \begin{minipage}[t]{0.49\linewidth}
    \centering
    \includegraphics[width=1\linewidth]{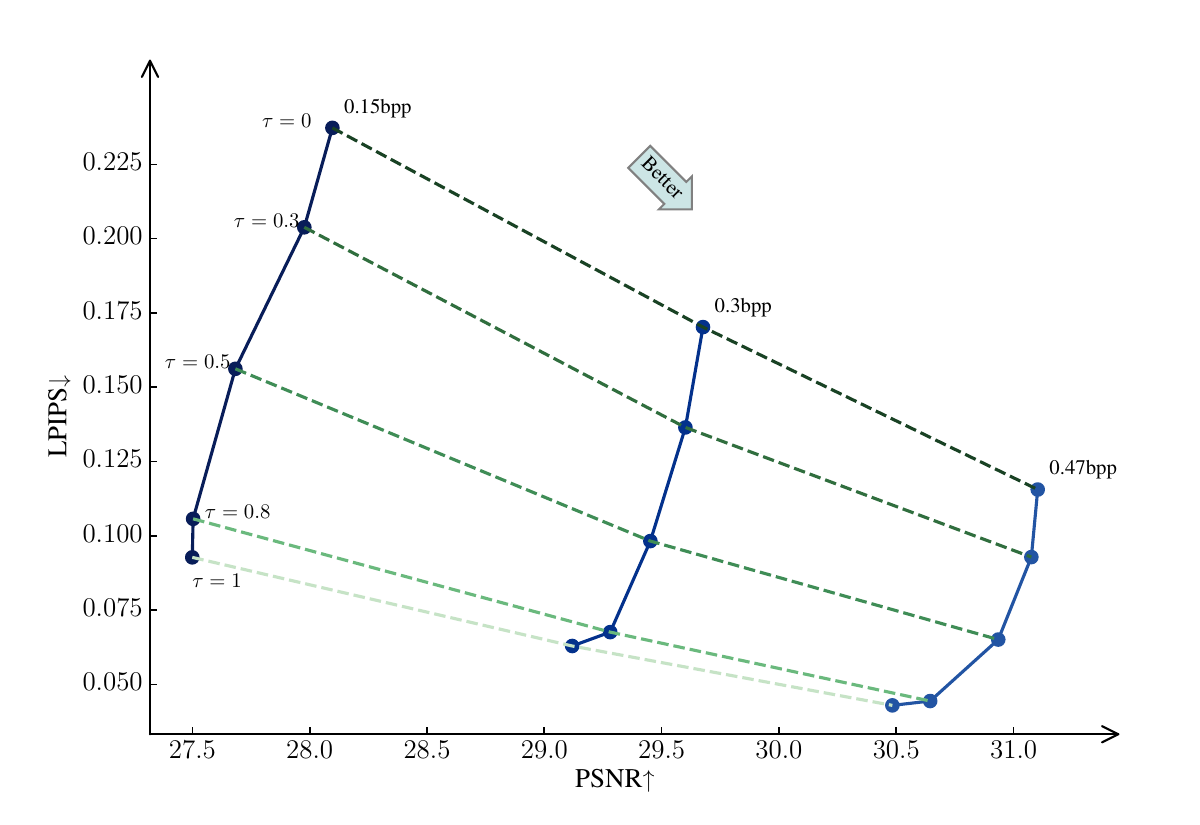}
    \end{minipage}%
    }%
    \centering
    \label{psnr_lpips}
    \caption{Distortion (PSNR) vs. perception(LPIPS) on Kodak for different rate-distortion-perception tradeoffs.}
\end{figure*}

\subsection{Visualization of features.}

\begin{figure}[htbp]
    \centering
    \includegraphics[width=1\linewidth]{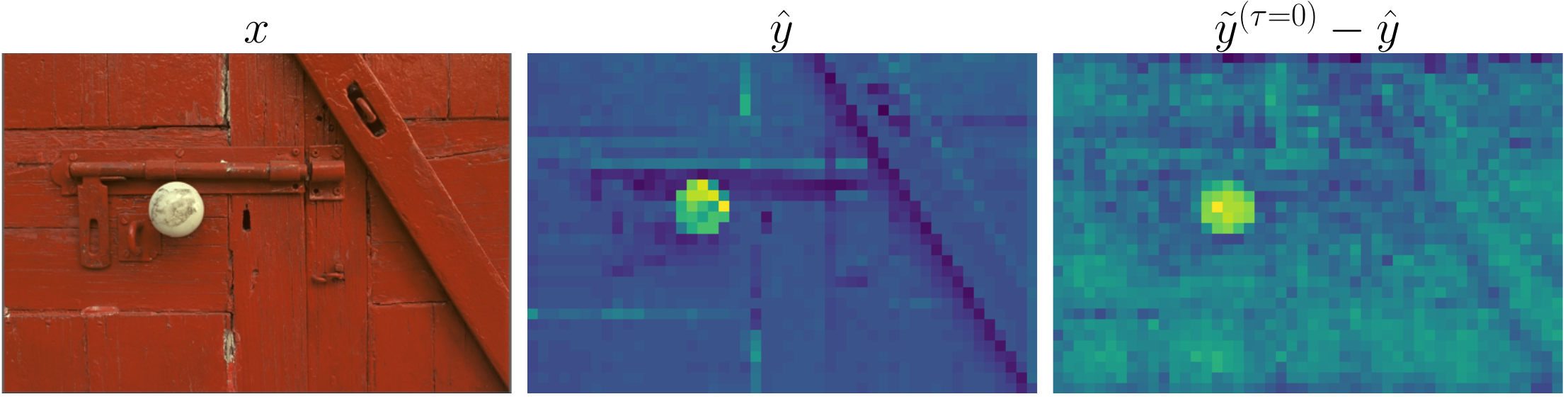}
    \caption{Visualization of original image $x$, compressed feature $\hat y$ and residual with transformed feature $\tilde y$. For better observation, feature and residual are normalized separately.}
    \label{res}
\end{figure}

We also visualize the residual between the transformed feature $\tilde y^{(\tau=0)}$ and the compressed feature $\hat y$ for distortion-oriented base model, as shown in the Fig. \ref{res}. $\tilde y$ introduces additional details, particularly enhancing the main objects in the original image $x$. 

\subsection{Diffusion Stability}

We conduct repeated experiments and observe stable results. In unconditional generation, diffusion may yield variable or failed outputs. However, our method's strong decoding condition constrains and stabilizes diffusion process, ensuring consistent results. Table \ref{tab:fluctuations} below shows PSNR and LPIPS variations over 10 runs on Ours(Cheng2020)$\tau=0$, with minor fluctuations.

\begin{table}[h]
\centering
\begin{tabular}{c|c|c}
\toprule[1pt]
Dataset & PSNR & LPIPS \\ 
\midrule[1pt]
Kodak (BPP=0.2686) & 30.3089 $\pm$ 0.0005 & 0.10122 $\pm$ 0.00008 \\
\bottomrule[1pt]
\end{tabular}
\caption{Fluctuations in PSNR and LPIPS for the Kodak dataset using Ours(Cheng2020)$\tau=0$.}
\label{tab:fluctuations}
\end{table}

\subsection{More Metrics}

In the paper, we already include LPIPS, FID, CLIP-IQA, and NIQE as perceptual metrics. We also test on more subjective metrics on Table \ref{tab:metrics} using Ours(Cheng2020), such as PIEAPP~\cite{Prashnani_18_CVPR_PIEAPP}, BRISQUE~\cite{Mittal_11_ACSCC_BRISQUE}, and PI~\cite{Blau_18_ECCV_PI}, which show that $\tau=0$ has a significant improvement in perception compared to base codec that is $\tau=1$.

\begin{table}[htbp]
\centering
\begin{tabular}{c|c|c|c}
\toprule[1pt]
\textbf{BPP, $\tau=0/\tau=1$} & \textbf{PIEAPP $\downarrow$} & \textbf{BRISQUE $\downarrow$} & \textbf{PI $\downarrow$} \\ 
\midrule[1pt]
0.1174 & 1.106 / 1.429 & 9.956 / 45.85 & 2.979 / 4.436 \\ 
0.2686 & 0.6300 / 0.8767 & 13.93 / 34.47 & 2.714 / 3.551 \\
0.5930 & 0.4003 / 0.5073 & 10.62 / 24.12 & 2.374 / 2.854 \\ 
\bottomrule[1pt]
\end{tabular}
\caption{Performance comparison of PIEAPP, BRISQUE, and PI metrics for $\tau=0$ and $\tau=1$.}
\label{tab:metrics}
\end{table}

\subsection{User study}
To assess human preference, we conduct a user study with 12 participants who evaluate 24 sets of images($\tau=0, 0.5, 1$) from Kodak, CLIC. Participants select the highest quality results (more realistic), with 93\% favoring processed images: 68\% prefer $\tau=0$, 25\% prefer $\tau=0.5$, and only 7\% favor $\tau=1$. These support effectiveness of our method and highlight the flexibility of controlling $\tau$. 

\subsection{Inference Latency}
Tested on a 4090 GPU with Kodak, our method shows substantial latency improvement compared with diffusion-based method. DiffBIR~\cite{Lin_2023_arxiv_Diffbir}, with DDIM acceleration, requires 50 diffusion steps, averaging 6953 ms per image. CDC~\cite{Yang_2023_NeurIPS_CDC} needs 17 image-domain steps, taking 1381 ms. Our method, performing diffusion in the feature domain with only 10 steps, requires just 331 ms. This shows that our decoding scheme has made great progress in latency compared to other diffusion-based methods. Our encoding does not introduce additional time and is the same as base codec.

\subsection{More Visualization}

\begin{figure*}[htbp]
    \centering
    \includegraphics[width=1\linewidth]{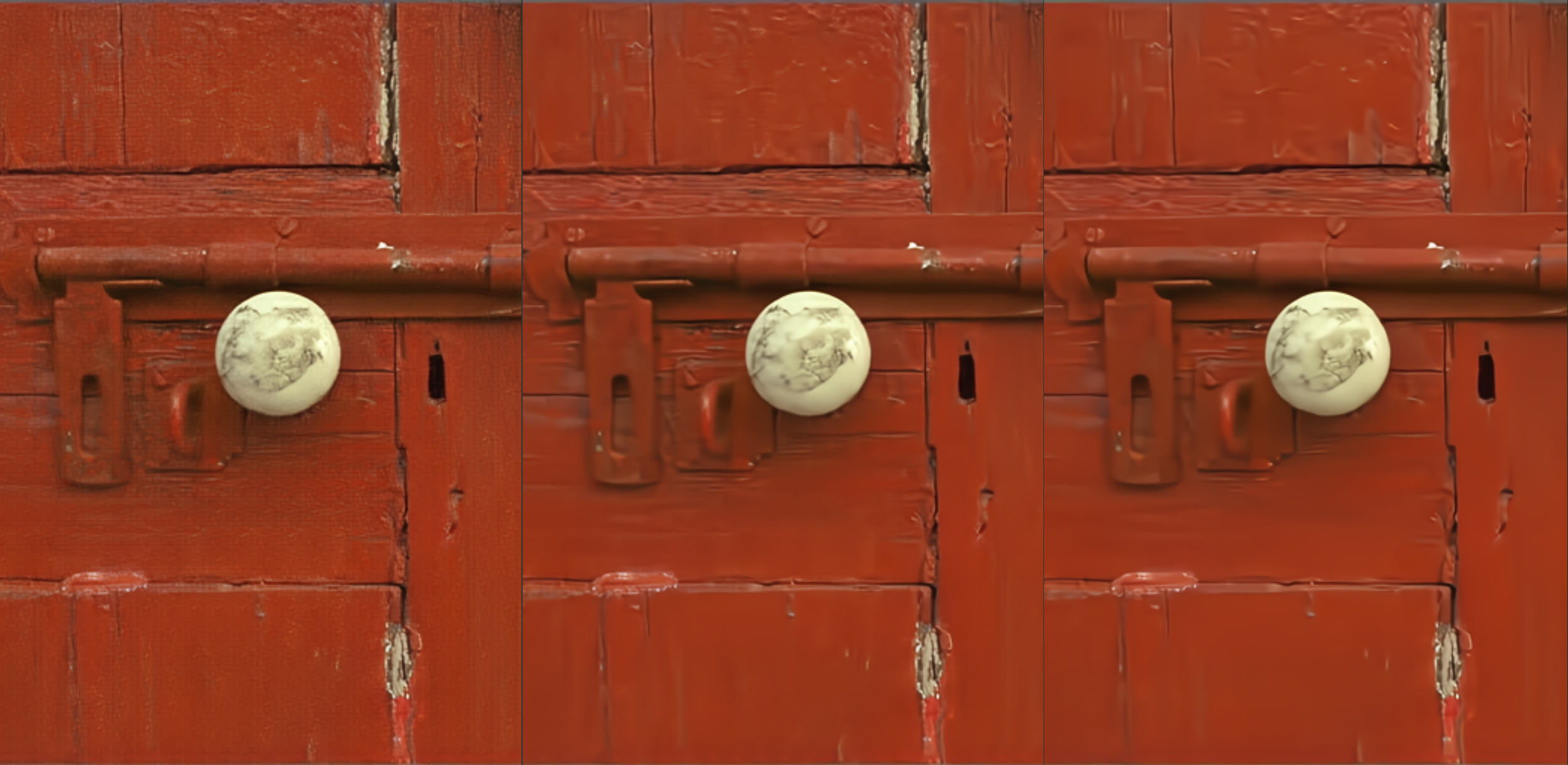}
    \caption{Reconstructions of Ours(Cheng2020) with different $\tau$. (Right) $\tau=0$, (Middle) $\tau=0.5$, (Left) $\tau=1$.}
    \label{fig:1}
\end{figure*}

\begin{figure*}[htbp]
    \centering
    \includegraphics[width=1\linewidth]{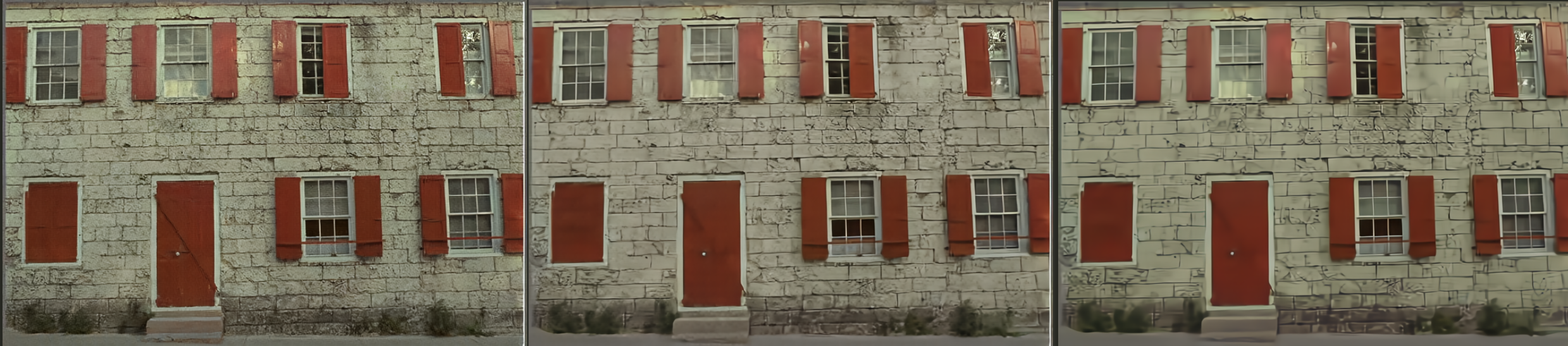}
    \caption{Reconstructions of Ours(Cheng2020) with different $\tau$. (Right) $\tau=0$, (Middle) $\tau=0.5$, (Left) $\tau=1$.}
    \label{fig:22}
\end{figure*}

\begin{figure*}[htbp]
    \centering
    \includegraphics[width=1\linewidth]{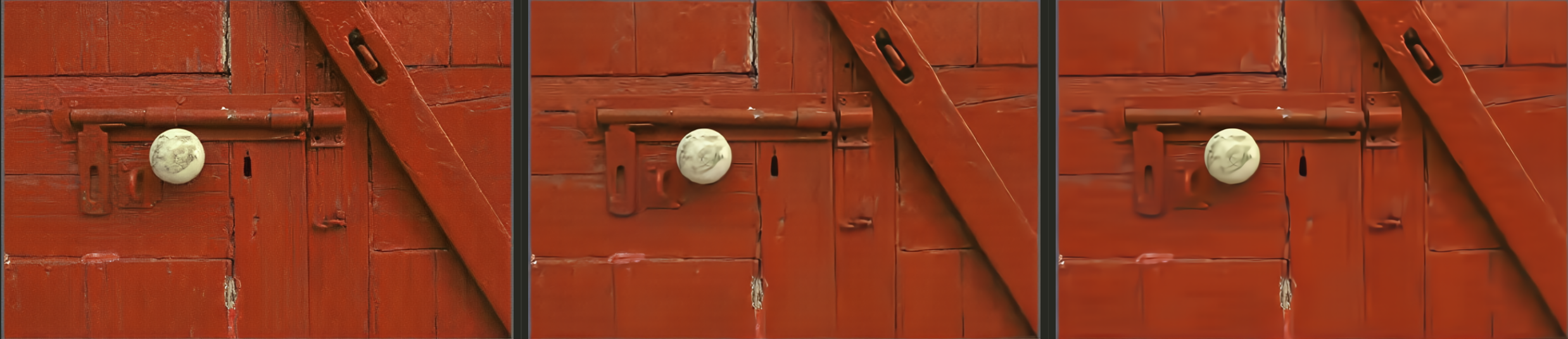}
    \caption{Reconstructions of Ours(Cheng2020) with different $\tau$. (Right) $\tau=0$, (Middle) $\tau=0.5$, (Left) $\tau=1$.}
    \label{fig:23}
\end{figure*}

\begin{figure*}[htbp]
    \centering
    \includegraphics[width=1\linewidth]{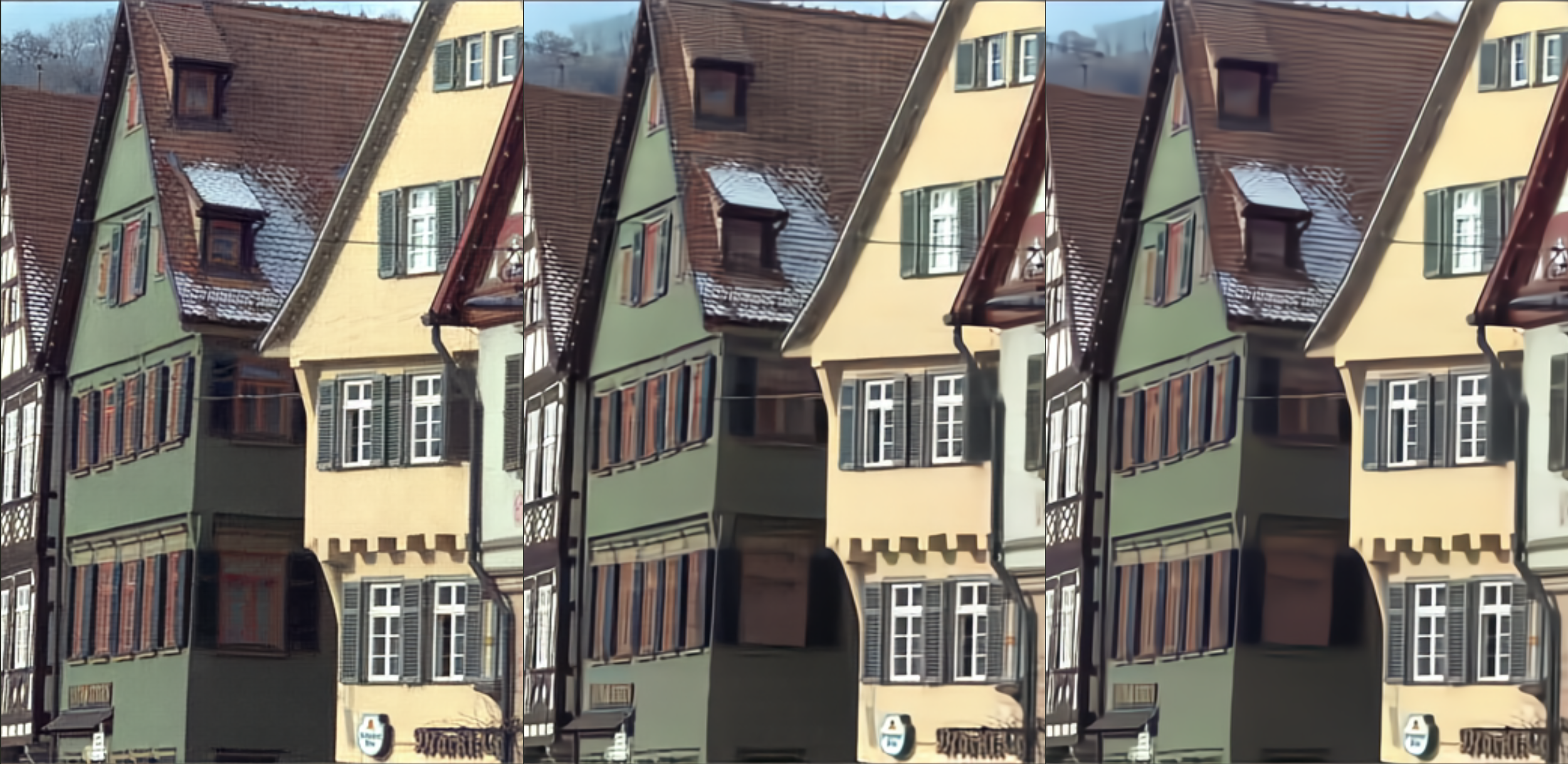}
    \caption{Reconstructions of Ours(Cheng2020) with different $\tau$. (Right) $\tau=0$, (Middle) $\tau=0.5$, (Left) $\tau=1$.}
    \label{fig:2}
\end{figure*}

\begin{figure*}[htbp]
    \centering
    \includegraphics[width=1\linewidth]{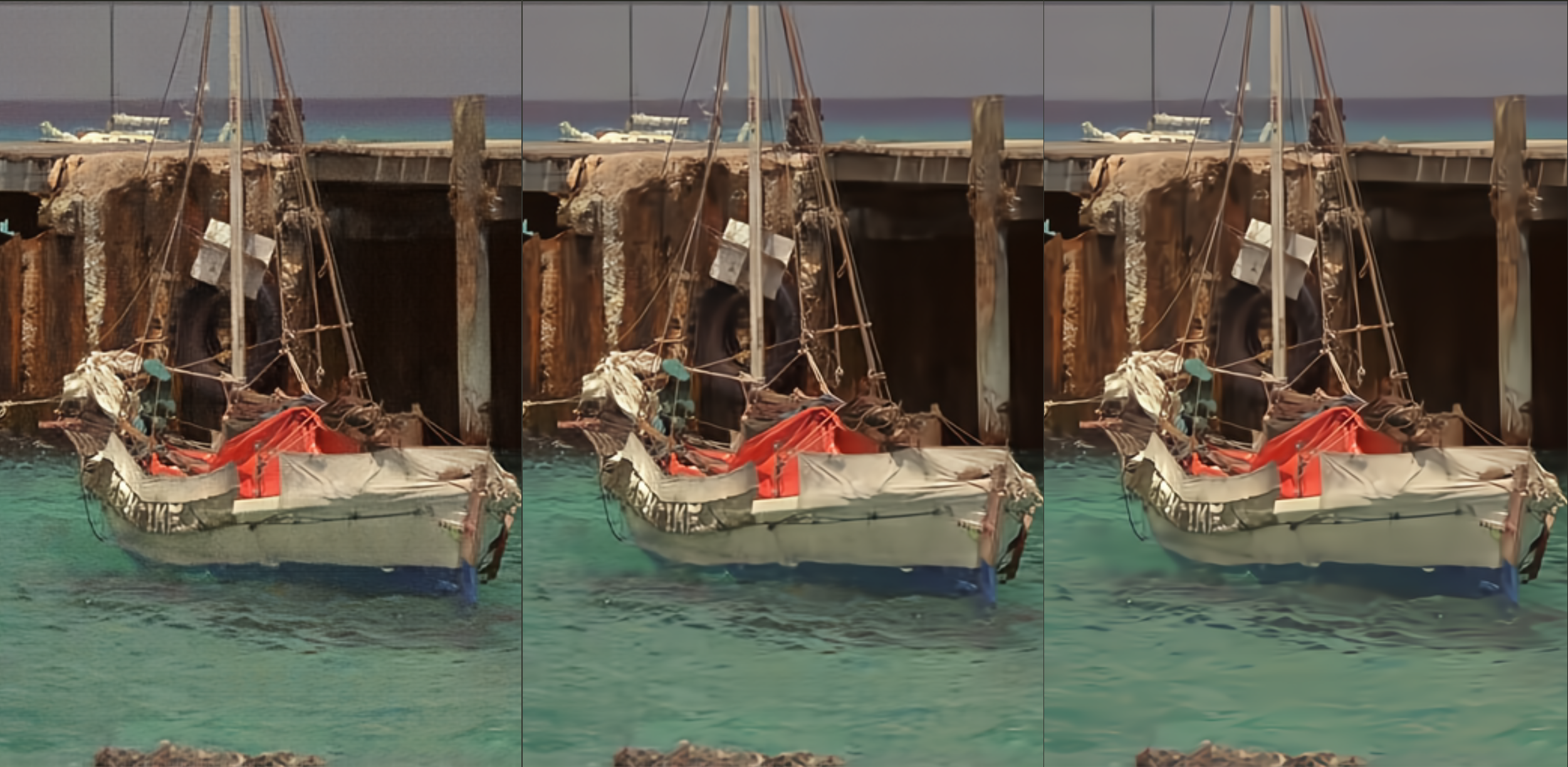}
    \caption{Reconstructions of Ours(Cheng2020) with different $\tau$. (Right) $\tau=0$, (Middle) $\tau=0.5$, (Left) $\tau=1$.}
    \label{fig:3}
\end{figure*}

\begin{figure*}[htbp]
    \centering
    \includegraphics[width=1\linewidth]{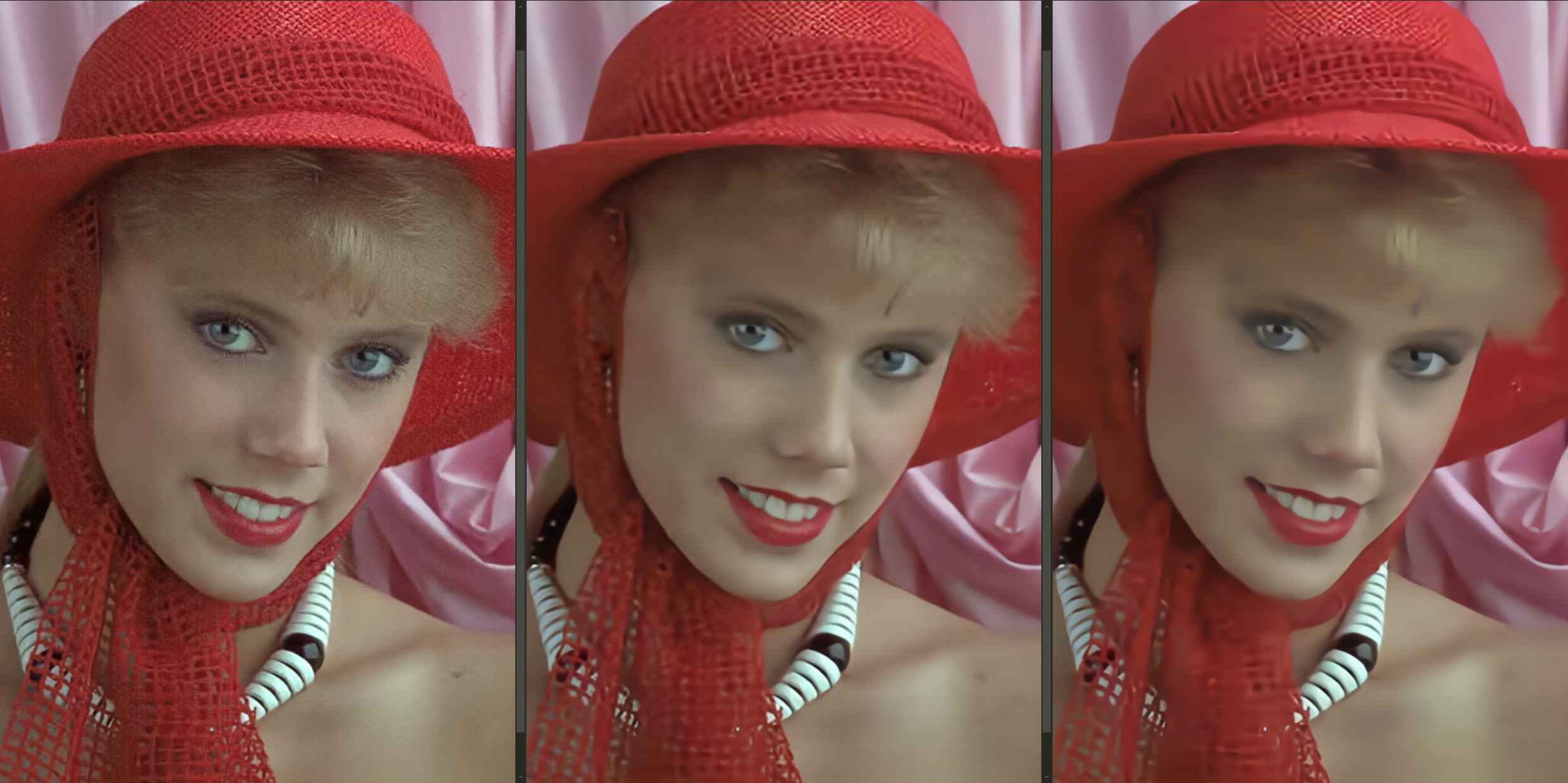}
    \caption{Reconstructions of Ours(Cheng2020) with different $\tau$. (Right) $\tau=0$, (Middle) $\tau=0.5$, (Left) $\tau=1$.}
    \label{fig:4}
\end{figure*}

\begin{figure*}[htbp]
    \centering
    \includegraphics[width=1\linewidth]{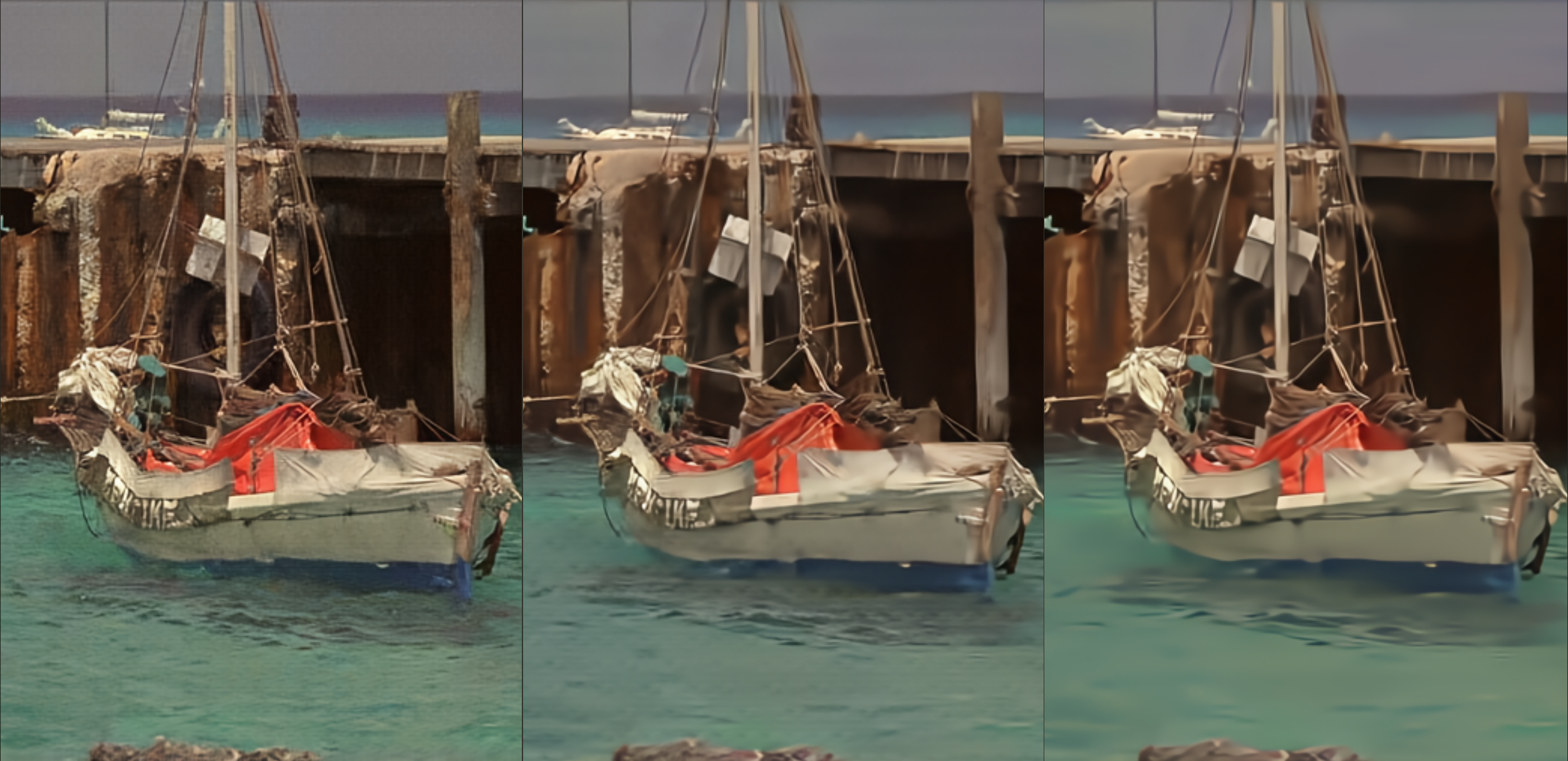}
    \caption{Reconstructions of Ours(Cheng2020) with different $\tau$. (Right) $\tau=0$, (Middle) $\tau=0.5$, (Left) $\tau=1$.}
    \label{fig:27}
\end{figure*}

\begin{figure*}[htbp]
    \centering
    \includegraphics[width=1\linewidth]{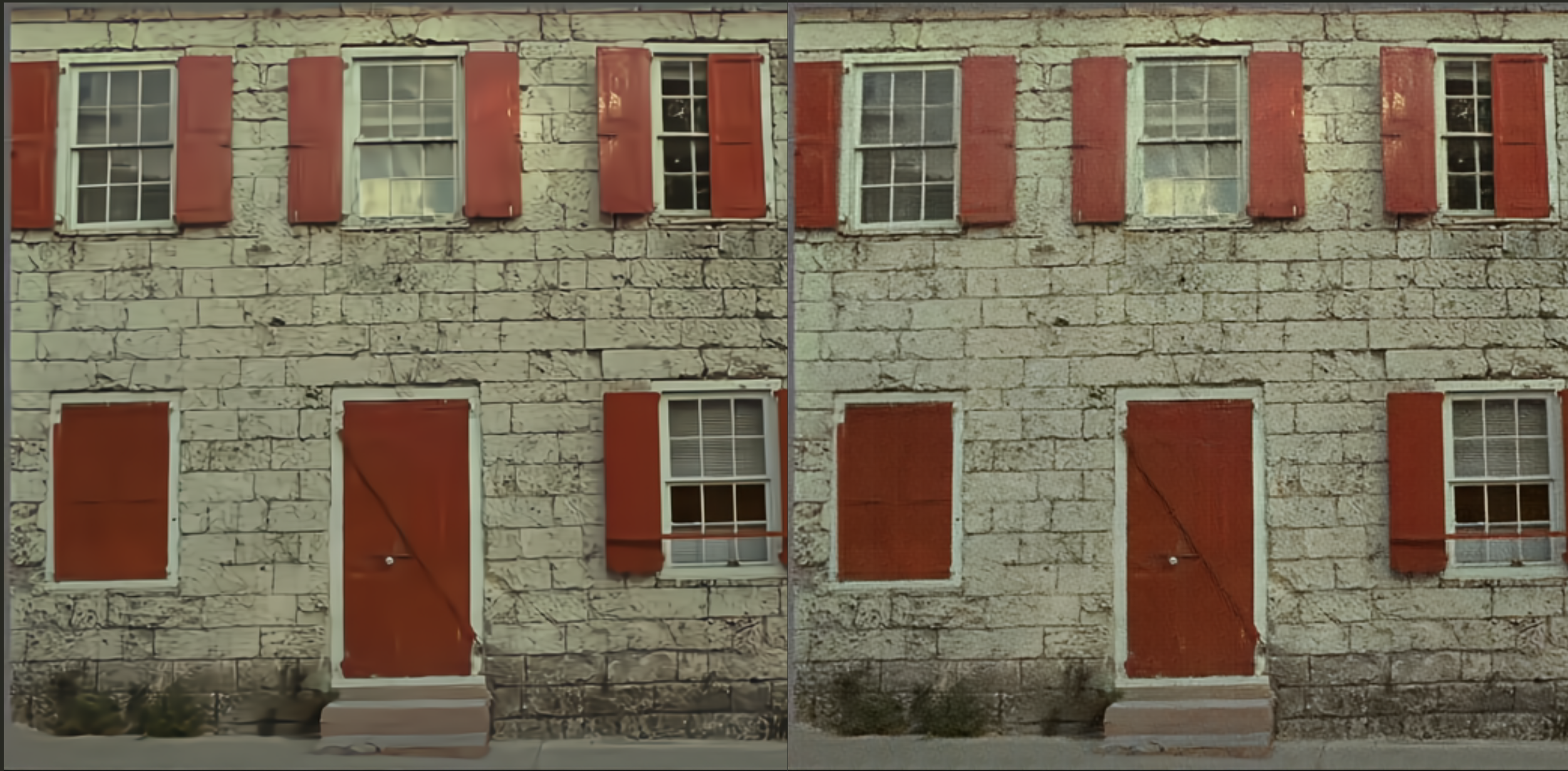}
    \caption{Reconstructions of Ours(Cheng2020) with different $\tau$. (Right) $\tau=1$, (Left) $\tau=0$.}
    \label{fig:5}
\end{figure*}

\begin{figure*}[htbp]
    \centering
    \includegraphics[width=1\linewidth]{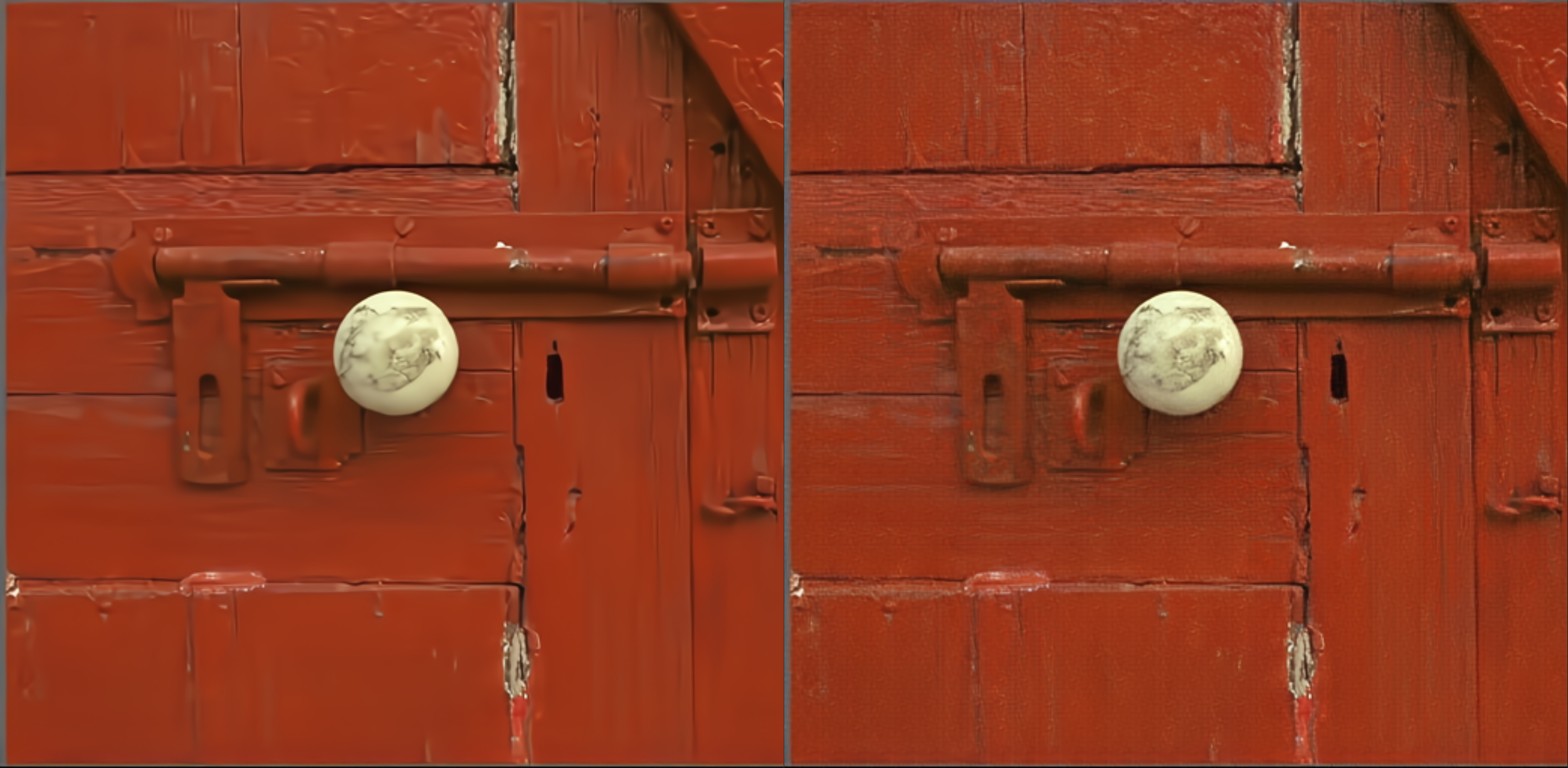}
    \caption{Reconstructions of Ours(Cheng2020) with different $\tau$. (Right) $\tau=1$, (Left) $\tau=0$.}
    \label{fig:6}
\end{figure*}

\begin{figure*}[htbp]
    \centering
    \includegraphics[width=1\linewidth]{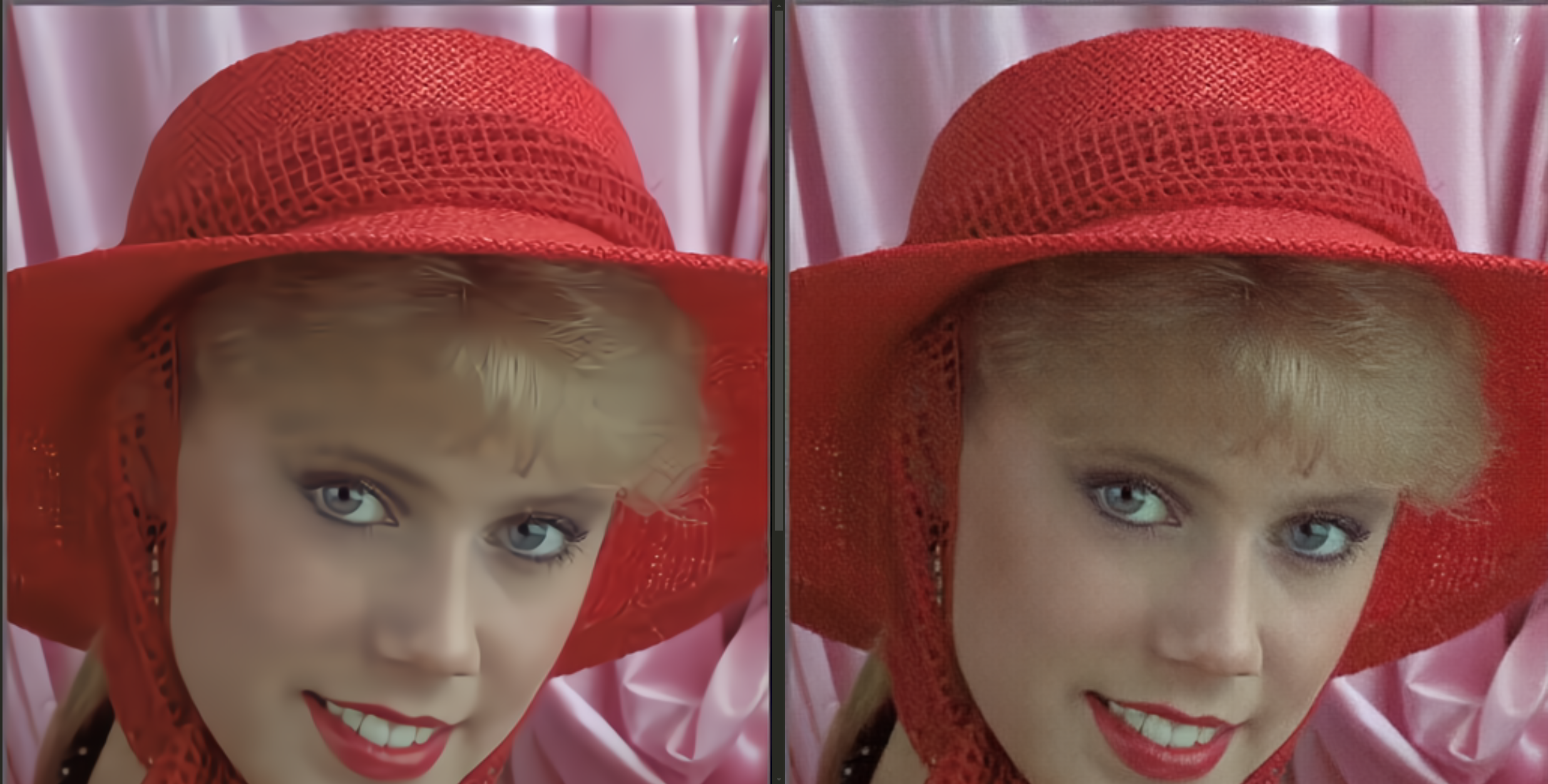}
    \caption{Reconstructions of Ours(Cheng2020) with different $\tau$. (Right) $\tau=1$, (Left) $\tau=0$.}
    \label{fig:7}
\end{figure*}

We present additional visual results of our method at different $\tau$ values, as illustrated in Fig. \ref{fig:1} to Fig. \ref{fig:7}. For Ours(Cheng2020), $\tau=0$ produces reconstructions with enhanced realism while using the same bit stream as $\tau=1$.

\end{document}